\documentclass{article}
\usepackage{amssymb,amsmath,amsfonts,graphicx,hyperref}
\newcommand{\mcm}[3]{\newcommand{#1}[#2]{{\ensuremath{#3}}}}

\usepackage{latexsym}
\usepackage{amssymb}

\mcm{\blank}{0}{(\emptybk)} \mcm{\dashbk}{0}{\mbox{---}}
\mcm{\emptybk}{0}{\:\:} \mcm{\hyph}{0}{\mbox{-}}

\mcm{\diagspace}{0}{\mbox{\hspace{2em}}}

\mcm{\cat}{1}{\mc{#1}} \mcm{\fcat}{1}{\mb{#1}}
\mcm{\mc}{1}{\mathcal{#1}} \mcm{\mr}{1}{\mathrm{#1}}
\mcm{\mi}{1}{\mathit{#1}} \mcm{\mb}{1}{\mathbf{#1}}
\mcm{\scat}{1}{\Bbb{#1}} \mcm{\twid}{1}{\widetilde{#1}}

\mcm{\elt}{0}{\in} \mcm{\sub}{0}{\,\subseteq\,}
\mcm{\such}{0}{\:|\:} \mcm{\without}{0}{\setminus}

\mcm{\atsr}{0}{\Box} \mcm{\eqv}{0}{\,\simeq\,}
\mcm{\iso}{0}{\,\cong\,}
\mcm{\of}{0}{\raisebox{0.2mm}{\ensuremath{\scriptstyle\circ}}}

\mcm{\bdry}{0}{\partial}

\mcm{\Bee}{0}{\cat{B}} \mcm{\Beep}{0}{\cat{B'}}
\mcm{\Eee}{0}{\cat{E}} \mcm{\Eeep}{0}{\cat{E'}}

\mcm{\Ess}{0}{\cat{S}} \mcm{\Tee}{0}{\cat{T}}
\mcm{\Teep}{0}{\cat{T'}} \mcm{\Stee}{0}{\scat{T}}
\mcm{\Steep}{0}{\scat{T'}}

\mcm{\blbk}{0}{\blank^{\blob}}
\mcm{\blob}{0}{\scriptscriptstyle{\bullet}}
\mcm{\stbk}{0}{\blank^{*}} \mcm{\ubl}{0}{{}^{\blob}}
\mcm{\ust}{0}{{}^{*}}

\mcm{\Cartpr}{0}{\pr{\Eee}{T}} \mcm{\Cartprp}{0}{\pr{\Eeep}{T'}}
\mcm{\Mnd}{0}{\triple{T}{\eta}{\mu}}
\mcm{\Zeropr}{0}{\pr{\Set}{\id}}

\mcm{\dopset}{0}{\ftrcat{\Delta^{\op}}{\Set}}
\mcm{\tropset}{0}{\ftrcat{\fcat{TR}^{\op}}{\Set}}

\mcm{\cod}{0}{\mr{cod}} \mcm{\dom}{0}{\mr{dom}}
\mcm{\End}{0}{\mr{End}} \mcm{\Hom}{0}{\mr{Hom}}
\mcm{\ob}{0}{\mr{ob}\,} \mcm{\op}{0}{\mr{op}}

\mcm{\comp}{0}{\mi{comp}} \mcm{\id}{0}{\mi{id}}
\mcm{\ids}{0}{\mi{ids}} \mcm{\mult}{0}{\mi{mult}}
\mcm{\unit}{0}{\mi{unit}}

\mcm{\Ab}{0}{\fcat{Ab}} \mcm{\Alg}{0}{\fcat{Alg}}
\mcm{\Bim}{1}{\fcat{Bim}(#1)} \mcm{\Cat}{0}{\fcat{Cat}}
\mcm{\Cay}{0}{\fcat{Cay}} \mcm{\Cpn}{1}{\pr{\Set/S_{#1}}{T_{#1}}}
\mcm{\fc}{0}{\fcat{fc}} \mcm{\fm}{0}{\fcat{fm}}
\mcm{\Graph}{0}{\fcat{Graph}} \mcm{\Gy}{0}{\fcat{Gy}}
\mcm{\Hpn}{1}{\pr{\Eee_{#1}}{P_{#1}}} \mcm{\Mon}{0}{\mb{Mon}}
\mcm{\Multicat}{0}{\fcat{Multicat}} \mcm{\One}{0}{\fcat{1}}
\mcm{\PD}{1}{\fcat{PD}_{#1}} \mcm{\Prof}{0}{\fcat{Prof}}
\mcm{\Set}{0}{\fcat{Set}} \mcm{\Span}{0}{\fcat{Span}}
\mcm{\Ssq}{0}{\fcat{Ssq}} \mcm{\Struc}{0}{\fcat{Struc}}
\mcm{\Sym}{0}{\fcat{Sym}} \mcm{\TR}{1}{\fcat{TR}(#1)}
\mcm{\Tr}{0}{\fcat{Tr}} \mcm{\Twocat}{0}{\fcat{2\hyph\Cat}}

\mcm{\integers}{0}{\mathbb{Z}}

\mcm{\range}{2}{#1,\,\ldots\,,#2}
\mcm{\bftuple}{2}{\tuplebts{\range{#1}{#2}}}
\mcm{\tuple}{3}{\tuplebts{\range{#1,#2}{#3}}}
\mcm{\rttuple}{1}{\tuplebts{\,\ldots\,,#1}}
\mcm{\abftuple}{2}{\atuplebts{\range{#1}{#2}}}
\mcm{\atuple}{3}{\atuplebts{\range{#1,#2}{#3}}}
\mcm{\arttuple}{1}{\atuplebts{\,\ldots\,,#1}}
\mcm{\sqbftuple}{2}{\obt\range{#1}{#2}\cbt}
\mcm{\pr}{2}{\tuplebts{#1,#2}}
\mcm{\triple}{3}{\tuplebts{#1,#2,#3}}

\mcm{\eend}{2}{#1[#2]} \mcm{\ehom}{3}{#1[#2,#3]}
\mcm{\ftrcat}{2}{[#1,#2]} \mcm{\homset}{3}{#1(#2,#3)}
\mcm{\multihom}{3}{#1(#2;#3)}
\mcm{\relhom}{5}{#1_{#2}(\range{#3}{#4};#5)}

\mcm{\go}{0}{\rTo} \mcm{\goby}{1}{\rTo^{#1}}
\mcm{\goesto}{0}{\,\longmapsto\,} \mcm{\goiso}{0}{\goby{\diso}}
\mcm{\monic}{0}{\rMonic} \mcm{\og}{0}{\lTo}
\mcm{\ogby}{1}{\lTo^{#1}}

\mcm{\gph}{2}{\spn{#1}{T #2}{#2}} \mcm{\graph}{4}{\spaan{#1}{T
#2}{#2}{#3}{#4}} \mcm{\oppair}{2}{\stackrel{\rTo^{#1}}{\lTo_{#2}}}
\mcm{\parpair}{2}{\stackrel{\rTo^{#1}}{\rTo_{#2}}}
\mcm{\spn}{3}{#2 \og #1 \go #3} \mcm{\spaan}{5}{#2 \ogby{#4} #1
\goby{#5} #3}

\mcm{\bktdvslob}{3}
    {\left(
    \begin{diagram}[height=1.5em]
    #1      \\
    \dTo>{\,#2} \\
    #3      \\
    \end{diagram}
    \right)}
\mcm{\slob}{3}{(#1 \goby{#2} #3)} \mcm{\vslob}{3}
    {\left.
    \begin{diagram}[height=1.5em]
    #1      \\
    \dTo>{\,#2} \\
    #3      \\
    \end{diagram}
    \right.}

\newenvironment{tree}
    {\begin{diagram}[height=1em,width=.75em,abut,noPS,tight]}
    {\end{diagram}}

\mcm{\enode}{0}{\circ}

\mcm{\nl}{1}{\stackrel{\textstyle #1}{\node}}
\mcm{\node}{0}{\bullet}

\mcm{\utree}{0}{\node}

\mcm{\diso}{0}{\sim}

\mcm{\vdiso}{0}{\wr}

\mcm{\nat}{0}{\mathbb{N}}

\mcm{\Onepr}{0}{\pr{\Graph}{\fc}}
\newlength{\nllwidth}
\newlength{\nllheight}
\newcommand{\stackbelow}[2]{%
\settowidth{\nllwidth}{\ensuremath{#1}\ensuremath{#2}}%
\settoheight{\nllheight}{\ensuremath{#2}}%
\addtolength{\nllheight}{.3ex}%
\mbox{%
\ensuremath{#1}%
\hspace{-.5\nllwidth}%
\raisebox{-1\nllheight}{\ensuremath{#2}}}}
\mcm{\nlal}{2}{\stackbelow{\nl{#1}}{#2}}
\mcm{\nll}{1}{\stackbelow{\node}{#1}} \mcm{\wun}{0}{\fcat{1}}
\mcm{\atuplebts}{1}{\langle #1 \rangle} \mcm{\tuplebts}{1}{(#1)}
\mcm{\bo}{0}{(} \mcm{\bc}{0}{)}
\mcm{\UBilax}{0}{\fcat{UBicat}_\mr{lax}}
\mcm{\UBiwk}{0}{\fcat{UBicat}_\mr{wk}}
\mcm{\UBistr}{0}{\fcat{UBicat}_\mr{str}}
\mcm{\Bilax}{0}{\fcat{Bicat}_\mr{lax}}
\mcm{\Biwk}{0}{\fcat{Bicat}_\mr{wk}}
\mcm{\Bistr}{0}{\fcat{Bicat}_\mr{str}} \mcm{\rotsub}{0}{\cup
\raisebox{0.1em}{$\scriptstyle{|}$}} \mcm{\pd}{0}{\fcat{pd}}
\mcm{\rep}{1}{\widehat{#1}} \mcm{\ovln}{1}{\overline{#1}}
\mcm{\Gph}{0}{\fcat{Gph}} \mcm{\tr}{0}{\fcat{tr}}

\mcm{\ladj}{0}{\,\dashv\,} \mcm{\zeropd}{0}{\node}
    {\end{diagram}}
\mcm{\END}{0}{\fcat{End}} \mcm{\HOM}{0}{\fcat{Hom}}

\newlength{\gwidth} 
\newlength{\gvert}  
\newlength{\gdrop}  
\newlength{\gbaredrop}  
\newlength{\goffset}    
\newlength{\gtemp}  

\newcommand{\present}[1]{%
\makebox[1\gwidth]{%
\rule[-1\gdrop]{0ex}{1\gvert}%
\raisebox{-1\gbaredrop}{#1}}}

\newcommand{\presentl}[1]{%
\makebox[1\gwidth][l]{%
\rule[-1\gdrop]{0ex}{1\gvert}%
\raisebox{-1\gbaredrop}{#1}}}

\newcommand{\presentr}[1]{%
\makebox[1\gwidth][r]{%
\rule[-1\gdrop]{0ex}{1\gvert}%
\raisebox{-1\gbaredrop}{#1}}}

\newcommand{\ginitdims}[2]{
\setlength{\unitlength}{1em}
\setlength{\goffset}{.25\unitlength}
\setlength{\gwidth}{#1\unitlength}
\setlength{\gvert}{#2\unitlength}
\setlength{\gdrop}{.5\gvert}
\addtolength{\gdrop}{-1\goffset}
\setlength{\gbaredrop}{1\gdrop}
\addtolength{\gvert}{.6\unitlength}
\addtolength{\gdrop}{.3\unitlength}}    

\newcommand{\cinitdims}[2]{
\setlength{\unitlength}{1em}
\setlength{\goffset}{.35\unitlength}
\setlength{\gwidth}{#1\unitlength}
\setlength{\gvert}{#2\unitlength}
\setlength{\gdrop}{.5\gvert}
\addtolength{\gdrop}{-1\goffset}
\setlength{\gbaredrop}{1\gdrop}
\addtolength{\gvert}{.6\unitlength}
\addtolength{\gdrop}{.3\unitlength}}    

\newcommand{\gsinitdims}[2]{
\setlength{\unitlength}{0.5em}
\setlength{\goffset}{.25\unitlength}
\setlength{\gwidth}{#1\unitlength}
\setlength{\gvert}{#2\unitlength}
\setlength{\gdrop}{.5\gvert}
\addtolength{\gdrop}{-1\goffset}
\setlength{\gbaredrop}{1\gdrop}
\addtolength{\gvert}{.6\unitlength}
\addtolength{\gdrop}{.3\unitlength}}    

\newcommand{\sidespic}[1]{%
\settowidth{\gtemp}{\ensuremath{#1}}%
\addtolength{\gwidth}{1\gtemp}}

\newcommand{\abovepic}[1]{%
\settoheight{\gtemp}{\ensuremath{#1}}%
\addtolength{\gvert}{1\gtemp}%
\settodepth{\gtemp}{\ensuremath{#1}}%
\addtolength{\gvert}{1\gtemp}}

\newcommand{\belowpic}[1]{%
\settoheight{\gtemp}{\ensuremath{#1}}%
\addtolength{\gvert}{1\gtemp}%
\addtolength{\gdrop}{1\gtemp}%
\settodepth{\gtemp}{\ensuremath{#1}}%
\addtolength{\gvert}{1\gtemp}%
\addtolength{\gdrop}{1\gtemp}}

\newcommand{\cell}[4]{\put(#1,#2){\makebox(0,0)[#3]{\ensuremath{#4}}}}
\mcm{\zmark}{0}{\scriptstyle{\bullet}}

\newcommand{\pregfst}[1]{%
\begin{picture}(0.5,0.2)(-0.5,-0.2)%
\cell{-0.1}{-0.2}{tr}{#1}%
\cell{0}{0}{c}{\zmark}%
\end{picture}}

\mcm{\gfst}{1}{%
\ginitdims{0.5}{0.4}%
\sidespic{#1}%
\belowpic{#1}%
\presentr{\pregfst{#1}}}

\newcommand{\preglst}[1]{%
\begin{picture}(0.5,0.2)(0,-0.2)%
\cell{0.1}{-0.2}{tl}{#1}%
\cell{0.05}{0}{c}{\zmark}%
\end{picture}}

\mcm{\glst}{1}{%
\ginitdims{.5}{.4}%
\sidespic{#1}%
\belowpic{#1}%
\presentl{\preglst{#1}}}

\newcommand{\preglft}[1]{%
\begin{picture}(0,0.2)(0,-0.2)%
\cell{-0.1}{-0.2}{tr}{#1}%
\cell{0.05}{0}{c}{\zmark}%
\end{picture}}

\mcm{\glft}{1}{%
\ginitdims{0}{.4}%
\belowpic{#1}%
\present{\preglft{#1}}}

\newcommand{\pregrgt}[1]{%
\begin{picture}(0,0.2)(0,-0.2)%
\cell{0.1}{-0.2}{tl}{#1}%
\cell{0.05}{0}{c}{\zmark}%
\end{picture}}

\mcm{\grgt}{1}{%
\ginitdims{0}{.4}%
\belowpic{#1}%
\present{\pregrgt{#1}}}

\newcommand{\pregblw}[1]{%
\begin{picture}(0,0.3)(0,-0.3)
\cell{0}{-0.3}{t}{#1}%
\cell{0.05}{0}{c}{\zmark}%
\end{picture}}

\mcm{\gblw}{1}{%
\ginitdims{0}{.6}%
\belowpic{#1}%
\present{\pregblw{#1}}}

\newcommand{\pregfbw}[1]{%
\begin{picture}(0,0.65)(0,-0.65)
\cell{0}{-0.65}{t}{#1}%
\cell{0.05}{0}{c}{\zmark}%
\end{picture}}

\mcm{\gfbw}{1}{%
\ginitdims{0}{1.3}%
\belowpic{#1}%
\present{\pregfbw{#1}}}

\newcommand{\pregzero}[1]{%
\begin{picture}(0.8,0.4)(-0.4,-0.4)
\cell{0}{-0.4}{t}{#1}%
\cell{0}{0}{c}{\zmark}%
\end{picture}}

\mcm{\gzero}{1}{%
\ginitdims{0.8}{.6}%
\belowpic{#1}%
\sidespic{#1}%
\present{\pregzero{#1}}}

\newcommand{\pregone}[1]{%
\begin{picture}(5,0.4)(0,-0.2)%
\cell{2.5}{0.2}{b}{#1}%
\put(0,0){\vector(1,0){5}}%
\end{picture}}

\mcm{\gone}{1}{%
\ginitdims{5}{0.4}%
\abovepic{#1}%
\present{\pregone{#1}}}

\newcommand{\pregtwo}[3]{%
\begin{picture}(5,3.4)(0,-0.2)%
\cell{2.5}{3.2}{b}{#1}%
\cell{2.5}{-.2}{t}{#2}%
\cell{2.7}{1.5}{l}{#3}%
\qbezier(0,1.5)(2.5,4.5)(5,1.5)%
\qbezier(0,1.5)(2.5,-1.5)(5,1.5)%
\put(5,1.5){\vector(1,-1){0}}%
\put(5,1.5){\vector(1,1){0}}%
\put(2.5,2.5){\vector(0,-1){2}}%
\end{picture}}

\mcm{\gtwo}{3}{%
\ginitdims{5}{3.4}%
\abovepic{#1}%
\belowpic{#2}%
\present{\pregtwo{#1}{#2}{#3}}}

\newcommand{\pregthree}[5]{%
\begin{picture}(5,5.4)(0,-1.2)%
\cell{2.5}{4.2}{b}{#1}%
\cell{1.5}{1.7}{b}{#2}%
\cell{2.5}{-1.2}{t}{#3}%
\cell{2.7}{2.75}{l}{#4}%
\cell{2.7}{0.25}{l}{#5}%
\qbezier(0,1.5)(2.5,6.5)(5,1.5)%
\qbezier(0,1.5)(2.5,-3.5)(5,1.5)%
\put(0,1.5){\vector(1,0){5}}%
\put(2.5,3.5){\vector(0,-1){1.5}}%
\put(2.5,1){\vector(0,-1){1.5}}%
\put(5,1.5){\vector(1,-3){0}}%
\put(5,1.5){\vector(1,3){0}}%
\end{picture}}

\mcm{\gthree}{5}{%
\ginitdims{5}{5.4}%
\abovepic{#1}%
\belowpic{#3}%
\present{\pregthree{#1}{#2}{#3}{#4}{#5}}}

\newcommand{\pregfour}[7]{%
\begin{picture}(5,8.4)(0,-2.7)%
\cell{2.5}{5.7}{b}{#1}%
\cell{1.5}{2.8}{b}{#2}%
\cell{1.5}{0.2}{t}{#3}%
\cell{2.5}{-2.7}{t}{#4}%
\cell{2.7}{4.25}{l}{#5}%
\cell{2.7}{1.5}{l}{#6}%
\cell{2.7}{-1.25}{l}{#7}%
\qbezier(0,1.5)(2.5,9.5)(5,1.5)%
\qbezier(0,1.5)(2.5,4)(5,1.5)%
\qbezier(0,1.5)(2.5,-1)(5,1.5)%
\qbezier(0,1.5)(2.5,-6.5)(5,1.5)%
\put(2.5,5.25){\vector(0,-1){2}}%
\put(2.5,2.5){\vector(0,-1){2}}%
\put(2.5,-0.25){\vector(0,-1){2}}%
\put(5,1.5){\vector(1,-4){0}}%
\put(5,1.5){\vector(4,-3){0}}%
\put(5,1.5){\vector(4,3){0}}%
\put(5,1.5){\vector(1,4){0}}%
\end{picture}}

\mcm{\gfour}{7}{%
\ginitdims{5}{8.4}%
\abovepic{#1}%
\belowpic{#4}%
\present{\pregfour{#1}{#2}{#3}{#4}{#5}{#6}{#7}}}

\newcommand{\pregthreecell}[5]{%
\begin{picture}(8,5)(-4,-2.5)%
\cell{0}{2.5}{b}{#1}%
\cell{0}{-2.5}{t}{#2}%
\cell{-1.7}{0}{r}{#3}%
\cell{1.7}{0}{l}{#4}%
\cell{0}{0.2}{b}{#5}%
\qbezier(-4,0)(0,4.2)(4,0)%
\qbezier(-4,0)(0,-4.2)(4,0)%
\qbezier(-0.5,1.8)(-2.5,0)(-0.5,-1.8)%
\qbezier(0.5,1.8)(2.5,0)(0.5,-1.8)%
\put(-1,0){\vector(1,0){2}}%
\put(4,0){\vector(1,-1){0}}%
\put(4,0){\vector(1,1){0}}%
\put(-0.5,-1.8){\vector(1,-1){0}}%
\put(0.5,-1.8){\vector(-1,-1){0}}%
\end{picture}}

\mcm{\gthreecell}{5}{%
\ginitdims{8}{5}%
\abovepic{#1}%
\belowpic{#2}%
\present{\pregthreecell{#1}{#2}{#3}{#4}{#5}}}

\newcommand{\pregthreecellu}{%
\begin{picture}(5,3.4)(-0.5,-0.2)%
\qbezier(-.5,1.5)(2,4.5)(4.5,1.5)%
\qbezier(-.5,1.5)(2,-1.5)(4.5,1.5)%
\qbezier(1.5,2.7)(0.5,1.5)(1.5,0.3)%
\qbezier(2.5,2.7)(3.5,1.5)(2.5,0.3)%
\put(1.3,1.5){\vector(1,0){1.4}}%
\put(4.5,1.5){\vector(1,-1){0}}%
\put(4.5,1.5){\vector(1,1){0}}%
\put(1.5,0.3){\vector(2,-3){0}}%
\put(2.5,0.3){\vector(-2,-3){0}}%
\end{picture}}

\mcm{\gthreecellu}{0}{%
\ginitdims{5}{3.4}%
\present{\pregthreecellu}}

\newcommand{\pregtwocentre}[3]{%
\begin{picture}(5,3.4)(0,-0.2)%
\cell{2.5}{3.2}{b}{#1}%
\cell{2.5}{-.2}{t}{#2}%
\cell{2.5}{1.5}{c}{#3}%
\qbezier(0,1.5)(2.5,4.5)(5,1.5)%
\qbezier(0,1.5)(2.5,-1.5)(5,1.5)%
\put(5,1.5){\vector(1,-1){0}}%
\put(5,1.5){\vector(1,1){0}}%
\put(2.5,2.5){\vector(0,-1){2}}%
\end{picture}}

\mcm{\gtwocentre}{3}{%
\ginitdims{5}{3.4}%
\abovepic{#1}%
\belowpic{#2}%
\present{\pregtwocentre{#1}{#2}{#3}}}

\newcommand{\pregspecialone}[9]{%
\begin{picture}(8,8)(-4,-4)%
\cell{0}{3.9}{b}{#1}%
\cell{-2}{-0.2}{t}{#2}%
\cell{0}{-3.9}{t}{#3}%
\cell{-1.5}{1.1}{r}{#4}%
\cell{0.2}{1.5}{l}{#5}%
\cell{1.5}{1.1}{l}{#6}%
\cell{0.2}{-2}{l}{#7}%
\cell{-0.9}{2.3}{b}{#8}%
\cell{0.9}{2.3}{b}{#9}%
\qbezier(-4,0)(0,8)(4,0)%
\qbezier(-4,0)(0,-8)(4,0)%
\qbezier(-0.5,3.4)(-3.5,2)(-0.5,0.6)%
\qbezier(0.5,3.4)(3.5,2)(0.5,0.6)%
\put(-4,0){\vector(1,0){8}}%
\put(0,3.4){\vector(0,-1){2.8}}%
\put(0,-0.8){\vector(0,-1){2.4}}%
\put(-1.5,2.2){\vector(1,0){1.2}}%
\put(0.3,2.2){\vector(1,0){1.2}}%
\put(4,0){\vector(1,-2){0}}%
\put(4,0){\vector(1,2){0}}%
\put(-0.5,0.6){\vector(2,-1){0}}%
\put(0.5,0.6){\vector(-2,-1){0}}%
\end{picture}}

\mcm{\gspecialone}{9}{%
\ginitdims{8}{8}%
\abovepic{#1}%
\belowpic{#3}%
\present{\pregspecialone{#1}{#2}{#3}{#4}{#5}{#6}{#7}{#8}{#9}}}

\newcommand{\pregspecialtwo}{%
\begin{picture}(5,3.4)(0,-0.2)%
\qbezier(0,1.5)(2.5,4.5)(5,1.5)%
\qbezier(0,1.5)(2.5,-1.5)(5,1.5)%
\qbezier(1.7,2.5)(0,1.5)(1.7,0.5)%
\qbezier(3.3,2.5)(5,1.5)(3.3,0.5)%
\put(5,1.5){\vector(1,-1){0}}%
\put(5,1.5){\vector(1,1){0}}%
\put(1.7,0.5){\vector(3,-2){0}}%
\put(3.3,0.5){\vector(-3,-2){0}}%
\put(2.5,2.5){\vector(0,-1){2}}%
\put(1.2,1.5){\vector(1,0){1}}%
\put(2.8,1.5){\vector(1,0){1}}%
\end{picture}}

\mcm{\gspecialtwo}{0}{%
\ginitdims{5}{3.4}%
\present{\pregspecialtwo}}

\newcommand{\pregspecialthree}{%
\begin{picture}(5,5.4)(0,-1.2)%
\qbezier(0,1.5)(2.5,6.5)(5,1.5)%
\qbezier(0,1.5)(2.5,-3.5)(5,1.5)%
\qbezier(2,3.5)(1,2.75)(2,2)%
\qbezier(3,3.5)(4,2.75)(3,2)%
\qbezier(2,1)(1,0.25)(2,-0.5)%
\qbezier(3,1)(4,0.25)(3,-0.5)%
\put(0,1.5){\vector(1,0){5}}%
\put(1.5,2.75){\vector(1,0){2}}%
\put(1.5,0.25){\vector(1,0){2}}%
\put(5,1.5){\vector(1,-3){0}}%
\put(5,1.5){\vector(1,3){0}}%
\put(2,2){\vector(1,-1){0}}%
\put(3,2){\vector(-1,-1){0}}%
\put(2,-0.5){\vector(1,-1){0}}%
\put(3,-0.5){\vector(-1,-1){0}}%
\end{picture}}

\mcm{\gspecialthree}{0}{%
\ginitdims{5}{5.4}%
\present{\pregspecialthree}}

\newcommand{\pregonew}[1]{%
\begin{picture}(8,0.4)(0,-0.2)%
\cell{4}{0.2}{b}{#1}%
\put(0,0){\vector(1,0){8}}%
\end{picture}}

\mcm{\gonew}{1}{%
\ginitdims{8}{0.4}%
\abovepic{#1}%
\present{\pregonew{#1}}}

\mcm{\gzersu}{0}{%
\gsinitdims{0}{.6}%
\present{\pregblw{}}}

\mcm{\gonesu}{0}{%
\gsinitdims{5}{0.4}%
\present{\pregone{}}}

\mcm{\gtwosu}{0}{%
\gsinitdims{5}{3.4}%
\present{\pregtwo{}{}{}}}

\mcm{\gthreesu}{0}{%
\gsinitdims{5}{5.4}%
\present{\pregthree{}{}{}{}{}}}

\mcm{\gfoursu}{0}{%
\gsinitdims{5}{8.4}%
\present{\pregfour{}{}{}{}{}{}{}}}

\newcommand{\precone}[1]{%
\begin{picture}(4.2,0.4)(-0.3,-0.2)%
\cell{1.8}{0.2}{b}{#1}%
\put(0,0){\vector(1,0){3.6}}%
\end{picture}}

\mcm{\cone}{1}{%
\cinitdims{4.2}{0.4}%
\abovepic{#1}%
\present{\precone{#1}}}

\mcm{\gfstsu}{0}{%
\gsinitdims{0.5}{0.4}%
\presentr{\pregfst{}}}

\mcm{\glstsu}{0}{%
\gsinitdims{0.5}{0.4}%
\presentl{\preglst{}}}

\newcommand{\prectwodbl}[3]%
{\begin{picture}(4.2,3.4)(-0.1,-0.2)%
\cell{2}{3.2}{b}{#1}%
\cell{2}{-0.2}{t}{#2}%
\cell{2.3}{1.5}{l}{#3}%
\qbezier(0,2)(2,4)(4,2)%
\qbezier(0,1)(2,-1)(4,1)%
\put(4,2){\vector(1,-1){0}}%
\put(4,1){\vector(1,1){0}}%
\put(1.9,2.5){\line(0,-1){1.8}}%
\put(2.1,2.5){\line(0,-1){1.8}}%
\cell{2.01}{0.4}{b}{\vee}%
\end{picture}}

\mcm{\ctwodbl}{3}{%
\cinitdims{4.2}{3.4}%
\abovepic{#1}%
\belowpic{#2}%
\present{\prectwodbl{#1}{#2}{#3}}}

\newcommand{\precthreedbl}[5]{%
\begin{picture}(4.2,5.4)(-0.1,-0.2)%
\cell{2}{5.2}{b}{#1}%
\cell{1}{2.7}{b}{#2}%
\cell{2}{-.2}{t}{#3}%
\cell{2.3}{3.75}{l}{#4}%
\cell{2.3}{1.25}{l}{#5}%
\qbezier(0,3)(2,7)(4,3)%
\qbezier(0,2)(2,-2)(4,2)%
\put(0,2.5){\vector(1,0){4}}%
\put(1.9,4.5){\line(0,-1){1.3}}%
\put(2.1,4.5){\line(0,-1){1.3}}%
\cell{2.01}{2.9}{b}{\vee}%
\put(1.9,2){\line(0,-1){1.3}}%
\put(2.1,2){\line(0,-1){1.3}}%
\cell{2.01}{0.4}{b}{\vee}%
\put(4,3){\vector(1,-3){0}}%
\put(4,2){\vector(1,3){0}}%
\end{picture}}

\mcm{\cthreedbl}{5}{%
\cinitdims{4.2}{5.4}%
\abovepic{#1}%
\belowpic{#3}%
\present{\precthreedbl{#1}{#2}{#3}{#4}{#5}}}

\newcommand{\precthreecelltrp}[5]{%
\begin{picture}(8.2,5)(-4.1,-2.5)%
\cell{0}{2.5}{b}{#1}%
\cell{0}{-2.5}{t}{#2}%
\cell{-1.8}{0}{r}{#3}%
\cell{1.8}{0}{l}{#4}%
\cell{0}{0.3}{b}{#5}%
\qbezier(-4,0.5)(0,4)(4,0.5)%
\qbezier(-4,-0.5)(0,-4)(4,-0.5)%
\qbezier(-0.6,2)(-2.6,0)(-0.6,-2)%
\qbezier(-0.4,2)(-2.4,0)(-0.5,-1.9)%
\cell{-0.6}{-2}{b}{\lrcorner}%
\qbezier(0.4,2)(2.4,0)(0.5,-1.9)%
\qbezier(0.6,2)(2.6,0)(0.6,-2)%
\cell{0.65}{-2}{b}{\llcorner}%
\put(-1,0.15){\line(1,0){1.7}}%
\put(-1,0){\line(1,0){2}}%
\put(-1,-0.15){\line(1,0){1.7}}%
\cell{1.15}{0}{r}{>}%
\put(4,0.5){\vector(1,-1){0}}%
\put(4,-0.5){\vector(1,1){0}}%
\end{picture}}

\mcm{\cthreecelltrp}{5}{%
\cinitdims{8.2}{5}%
\abovepic{#1}%
\belowpic{#2}%
\present{\precthreecelltrp{#1}{#2}{#3}{#4}{#5}}}

\newcommand{\prectwo}[3]%
{\begin{picture}(4.2,3.4)(-0.1,-0.2)%
\cell{2}{3.2}{b}{#1}%
\cell{2}{-0.2}{t}{#2}%
\cell{2.2}{1.5}{l}{#3}%
\qbezier(0,2)(2,4)(4,2)%
\qbezier(0,1)(2,-1)(4,1)%
\put(4,2){\vector(1,-1){0}}%
\put(4,1){\vector(1,1){0}}%
\put(2,2.5){\vector(0,-1){2}}%
\end{picture}}

\mcm{\ctwo}{3}{%
\cinitdims{4.2}{3.4}%
\abovepic{#1}%
\belowpic{#2}%
\present{\prectwo{#1}{#2}{#3}}}

\newcommand{\precthree}[5]{%
\begin{picture}(4.2,5.4)(-0.1,-0.2)%
\cell{2}{5.2}{b}{#1}%
\cell{1}{2.7}{b}{#2}%
\cell{2}{-.2}{t}{#3}%
\cell{2.2}{3.75}{l}{#4}%
\cell{2.2}{1.25}{l}{#5}%
\qbezier(0,3)(2,7)(4,3)%
\qbezier(0,2)(2,-2)(4,2)%
\put(0,2.5){\vector(1,0){4}}%
\put(2,4.5){\vector(0,-1){1.5}}%
\put(2,2){\vector(0,-1){1.5}}%
\put(4,3){\vector(1,-3){0}}%
\put(4,2){\vector(1,3){0}}%
\end{picture}}

\mcm{\cthree}{5}{%
\cinitdims{4.2}{5.4}%
\abovepic{#1}%
\belowpic{#3}%
\present{\precthree{#1}{#2}{#3}{#4}{#5}}}

\newcommand{\prectwoop}[3]%
{\begin{picture}(4.2,3.4)(-0.1,-0.2)%
\cell{2}{3.2}{b}{#1}%
\cell{2}{-0.2}{t}{#2}%
\cell{2.2}{1.5}{l}{#3}%
\qbezier(0,2)(2,4)(4,2)%
\qbezier(0,1)(2,-1)(4,1)%
\put(0,2){\vector(-1,-1){0}}%
\put(0,1){\vector(-1,1){0}}%
\put(2,2.5){\vector(0,-1){2}}%
\end{picture}}

\mcm{\ctwoop}{3}{%
\cinitdims{4.2}{3.4}%
\abovepic{#1}%
\belowpic{#2}%
\present{\prectwoop{#1}{#2}{#3}}}

\newcommand{\prectwopar}[4]{%
\begin{picture}(4.2,3.4)(-0.1,-0.2)%
\cell{2}{3.2}{b}{#1}%
\cell{2}{-0.2}{t}{#2}%
\cell{1.6}{1.5}{r}{#3}%
\cell{2.4}{1.5}{l}{#4}%
\qbezier(0,2)(2,4)(4,2)%
\qbezier(0,1)(2,-1)(4,1)%
\put(4,2){\vector(1,-1){0}}%
\put(4,1){\vector(1,1){0}}%
\put(1.8,2.5){\vector(0,-1){2}}%
\put(2.2,2.5){\vector(0,-1){2}}%
\end{picture}}

\mcm{\ctwopar}{4}{%
\cinitdims{4.2}{3.4}%
\abovepic{#1}%
\belowpic{#2}%
\present{\prectwopar{#1}{#2}{#3}{#4}}}

\newcommand{\precthreein}[5]{%
\begin{picture}(4.2,5.4)(-0.1,-0.2)%
\cell{2}{5.2}{b}{#1}%
\cell{1}{2.7}{b}{#2}%
\cell{2}{-.2}{t}{#3}%
\cell{2.2}{3.75}{l}{#4}%
\cell{2.2}{1.25}{l}{#5}%
\qbezier(0,3)(2,7)(4,3)%
\qbezier(0,2)(2,-2)(4,2)%
\put(0,2.5){\vector(1,0){4}}%
\put(2,4.5){\vector(0,-1){1.5}}%
\put(2,0.5){\vector(0,1){1.5}}%
\put(4,3){\vector(1,-3){0}}%
\put(4,2){\vector(1,3){0}}%
\end{picture}}

\mcm{\cthreein}{5}{%
\cinitdims{4.2}{5.4}%
\abovepic{#1}%
\belowpic{#3}%
\present{\precthreein{#1}{#2}{#3}{#4}{#5}}}

\newcommand{\precthreecell}[5]{%
\begin{picture}(8.2,5)(-4.1,-2.5)%
\cell{0}{2.5}{b}{#1}%
\cell{0}{-2.5}{t}{#2}%
\cell{-1.7}{0}{r}{#3}%
\cell{1.7}{0}{l}{#4}%
\cell{0}{0.2}{b}{#5}%
\qbezier(-4,0.5)(0,4)(4,0.5)%
\qbezier(-4,-0.5)(0,-4)(4,-0.5)%
\qbezier(-0.5,2)(-2.5,0)(-0.5,-2)%
\qbezier(0.5,2)(2.5,0)(0.5,-2)%
\put(-1,0){\vector(1,0){2}}%
\put(4,0.5){\vector(1,-1){0}}%
\put(4,-0.5){\vector(1,1){0}}%
\put(-0.5,-2){\vector(1,-1){0}}%
\put(0.5,-2){\vector(-1,-1){0}}%
\end{picture}}

\mcm{\cthreecell}{5}{%
\cinitdims{8.2}{5}%
\abovepic{#1}%
\belowpic{#2}%
\present{\precthreecell{#1}{#2}{#3}{#4}{#5}}}

\newcommand{\precthreecellpar}[6]{%
\begin{picture}(8.2,5)(-4.1,-2.5)%
\cell{0}{2.5}{b}{#1}%
\cell{0}{-2.5}{t}{#2}%
\cell{-1.7}{0}{r}{#3}%
\cell{1.7}{0}{l}{#4}%
\cell{0}{0.4}{b}{#5}%
\cell{0}{-0.4}{t}{#6}%
\qbezier(-4,0.5)(0,4)(4,0.5)%
\qbezier(-4,-0.5)(0,-4)(4,-0.5)%
\qbezier(-0.5,2)(-2.5,0)(-0.5,-2)%
\qbezier(0.5,2)(2.5,0)(0.5,-2)%
\put(-1,0.2){\vector(1,0){2}}%
\put(-1,-0.2){\vector(1,0){2}}%
\put(4,0.5){\vector(1,-1){0}}%
\put(4,-0.5){\vector(1,1){0}}%
\put(-0.5,-2){\vector(1,-1){0}}%
\put(0.5,-2){\vector(-1,-1){0}}%
\end{picture}}

\mcm{\cthreecellpar}{6}{%
\cinitdims{8.2}{5}%
\abovepic{#1}%
\belowpic{#2}%
\present{\precthreecellpar{#1}{#2}{#3}{#4}{#5}{#6}}}

\newcommand{\prectwov}[5]{%
\begin{picture}(3.4,4.2)(0.8,0.9)%
\cell{2.5}{5.1}{b}{#1}%
\cell{2.5}{0.9}{t}{#2}%
\cell{0.8}{3}{r}{#3}%
\cell{4.2}{3}{l}{#4}%
\cell{2.5}{3.2}{b}{#5}%
\qbezier(2,5)(0,3)(2,1)%
\qbezier(3,5)(5,3)(3,1)%
\put(2,1){\vector(1,-1){0}}%
\put(3,1){\vector(-1,-1){0}}%
\put(1.5,3){\vector(1,0){2}}%
\end{picture}}

\mcm{\ctwov}{5}{%
\cinitdims{3.4}{4.2}%
\abovepic{#1}%
\belowpic{#2}%
\sidespic{#3}%
\sidespic{#4}%
\present{\prectwov{#1}{#2}{#3}{#4}{#5}}}

\newcommand{\precthreecellv}[7]{%
\begin{picture}(5,8.2)(0.5,-1.6)%
\cell{3}{6.6}{b}{#1}%
\cell{3}{-1.6}{t}{#2}%
\cell{0.5}{2.5}{r}{#3}%
\cell{5.5}{2.5}{l}{#4}%
\cell{3}{4.2}{b}{#5}%
\cell{3}{0.8}{t}{#6}%
\cell{3.2}{2.5}{l}{#7}%
\qbezier(3.5,6.5)(7,2.5)(3.5,-1.5)%
\qbezier(2.5,6.5)(-1,2.5)(2.5,-1.5)%
\put(2.5,-1.5){\vector(1,-1){0}}%
\put(3.5,-1.5){\vector(-1,-1){0}}%
\qbezier(1,3)(3,5)(5,3)%
\qbezier(1,2)(3,0)(5,2)%
\put(5,3){\vector(1,-1){0}}%
\put(5,2){\vector(1,1){0}}%
\put(3,3.5){\vector(0,-1){2}}%
\end{picture}}

\mcm{\cthreecellv}{7}{%
\cinitdims{5}{8.2}%
\abovepic{#1}%
\belowpic{#2}%
\sidespic{#3}%
\sidespic{#4}%
\present{\precthreecellv{#1}{#2}{#3}{#4}{#5}{#6}{#7}}}

\newcommand{\pretopez}[2]{%
\begin{picture}(2.6,2.3)(-1.3,-2.2)%
\cell{0}{-2.2}{t}{#1}%
\cell{0}{-1.2}{c}{#2}%
\qbezier(0,0)(-2,-2)(0,-2)%
\qbezier(0,0)(2,-2)(0,-2)%
\put(0,0){\vector(-1,1){0}}%
\end{picture}}

\mcm{\topez}{2}{%
\ginitdims{2.6}{2.3}%
\belowpic{#1}%
\present{\pretopez{#1}{#2}}}

\newcommand{\pretopea}[3]{%
\begin{picture}(4,1.9)(-2,-0,2)%
\cell{0}{1.7}{b}{#1}%
\cell{0}{-0.2}{t}{#2}%
\cell{0}{0.7}{c}{#3}%
\qbezier(-2,0)(0,3)(2,0)%
\put(-2,0){\vector(1,0){4}}%
\put(2,0){\vector(2,-3){0}}%
\end{picture}}

\mcm{\topea}{3}{%
\ginitdims{4}{1.9}%
\abovepic{#1}%
\belowpic{#2}%
\present{\pretopea{#1}{#2}{#3}}}

\newcommand{\pretopeb}[4]{%
\begin{picture}(4,2.2)(-2,-0.2)%
\cell{-1.1}{1}{br}{#1}%
\cell{1.1}{1}{bl}{#2}%
\cell{0}{-0.2}{t}{#3}%
\cell{0}{0.8}{c}{#4}%
\put(-2,0){\vector(1,1){2}}%
\put(0,2){\vector(1,-1){2}}%
\put(-2,0){\vector(1,0){4}}%
\end{picture}}

\mcm{\topeb}{4}{%
\ginitdims{4}{2.2}%
\belowpic{#3}%
\present{\pretopeb{#1}{#2}{#3}{#4}}}

\newcommand{\pretopec}[5]{%
\begin{picture}(4,2.2)(-2,-0.2)%
\cell{-1.8}{1}{br}{#1}%
\cell{0}{2.2}{b}{#2}%
\cell{1.8}{1}{bl}{#3}%
\cell{0}{-0.2}{t}{#4}%
\cell{0}{0.8}{c}{#5}%
\put(-2,0){\vector(1,2){1}}%
\put(-1,2){\vector(1,0){2}}%
\put(1,2){\vector(1,-2){1}}%
\put(-2,0){\vector(1,0){4}}%
\end{picture}}

\mcm{\topec}{5}{%
\ginitdims{4}{2.2}%
\sidespic{#1}%
\abovepic{#2}%
\sidespic{#3}%
\belowpic{#4}%
\present{\pretopec{#1}{#2}{#3}{#4}{#5}}}

\newcommand{\pretoped}[6]{%
\begin{picture}(4,2.5)(-2,-0.2)%
\cell{-2}{0.6}{br}{#1}%
\cell{-0.7}{2.2}{br}{#2}%
\cell{0.7}{2.2}{bl}{#3}%
\cell{2}{0.6}{bl}{#4}%
\cell{0}{-0.2}{t}{#5}%
\cell{0}{0.8}{c}{#6}%
\put(-2,0){\vector(1,3){0.5}}%
\put(-1.5,1.5){\vector(3,2){1.5}}%
\put(0,2.5){\vector(3,-2){1.5}}%
\put(1.5,1.5){\vector(1,-3){0.5}}%
\put(-2,0){\vector(1,0){4}}%
\end{picture}}

\mcm{\toped}{6}{%
\ginitdims{4}{2.5}%
\sidespic{#1}%
\abovepic{#2}%
\abovepic{#3}%
\sidespic{#4}%
\belowpic{#5}%
\present{\pretoped{#1}{#2}{#3}{#4}{#5}{#6}}}

\newcommand{\pretopeq}[5]{%
\begin{picture}(4,2.5)(-2,-0.2)%
\cell{-2}{0.6}{br}{#1}%
\cell{-1}{2.2}{br}{#2}%
\cell{2}{0.6}{bl}{#3}%
\cell{0}{-0.2}{t}{#4}%
\cell{0}{0.8}{c}{#5}%
\put(-2,0){\vector(1,3){0.5}}%
\put(-1.5,1.5){\vector(1,1){1}}%
\cell{0.9}{2.3}{c}{\ddots}
\put(1.5,1.5){\vector(1,-3){0.5}}%
\put(-2,0){\vector(1,0){4}}%
\end{picture}}

\mcm{\topeq}{5}{%
\ginitdims{4}{2.5}%
\sidespic{#1}%
\abovepic{#2}%
\sidespic{#3}%
\belowpic{#4}%
\present{\pretopeq{#1}{#2}{#3}{#4}{#5}}}

\newcommand{\pretopebase}[1]{%
\begin{picture}(4,0.4)(0,-0.2)%
\cell{2}{0.2}{b}{#1}%
\put(0,0){\vector(1,0){4}}%
\end{picture}}

\mcm{\topebase}{1}{%
\ginitdims{4}{0.4}%
\abovepic{#1}%
\present{\pretopebase{#1}}}

\newcommand{\pretopezs}[2]{%
\begin{picture}(2.6,2.3)(-1.3,-2.2)%
\cell{0}{-2.2}{t}{#1}%
\cell{0}{-1.2}{c}{#2}%
\qbezier(0,0)(-2,-2)(0,-2)%
\qbezier(0,0)(2,-2)(0,-2)%
\end{picture}}

\mcm{\topezs}{2}{%
\ginitdims{2.6}{2.3}%
\belowpic{#1}%
\present{\pretopezs{#1}{#2}}}

\newcommand{\pretopeas}[3]{%
\begin{picture}(4,1.9)(-2,-0,2)%
\cell{0}{1.7}{b}{#1}%
\cell{0}{-0.2}{t}{#2}%
\cell{0}{0.7}{c}{#3}%
\qbezier(-2,0)(0,3)(2,0)%
\put(-2,0){\line(1,0){4}}%
\end{picture}}

\mcm{\topeas}{3}{%
\ginitdims{4}{1.9}%
\abovepic{#1}%
\belowpic{#2}%
\present{\pretopeas{#1}{#2}{#3}}}

\newcommand{\pretopebs}[4]{%
\begin{picture}(4,2.2)(-2,-0.2)%
\cell{-1.1}{1}{br}{#1}%
\cell{1.1}{1}{bl}{#2}%
\cell{0}{-0.2}{t}{#3}%
\cell{0}{0.8}{c}{#4}%
\put(-2,0){\line(1,1){2}}%
\put(0,2){\line(1,-1){2}}%
\put(-2,0){\line(1,0){4}}%
\end{picture}}

\mcm{\topebs}{4}{%
\ginitdims{4}{2.2}%
\belowpic{#3}%
\present{\pretopebs{#1}{#2}{#3}{#4}}}

\newcommand{\pretopecs}[5]{%
\begin{picture}(4,2.2)(-2,-0.2)%
\cell{-1.8}{1}{br}{#1}%
\cell{0}{2.2}{b}{#2}%
\cell{1.8}{1}{bl}{#3}%
\cell{0}{-0.2}{t}{#4}%
\cell{0}{0.8}{c}{#5}%
\put(-2,0){\line(1,2){1}}%
\put(-1,2){\line(1,0){2}}%
\put(1,2){\line(1,-2){1}}%
\put(-2,0){\line(1,0){4}}%
\end{picture}}

\mcm{\topecs}{5}{%
\ginitdims{4}{2.2}%
\sidespic{#1}%
\abovepic{#2}%
\sidespic{#3}%
\belowpic{#4}%
\present{\pretopecs{#1}{#2}{#3}{#4}{#5}}}

\newcommand{\pretopeds}[6]{%
\begin{picture}(4,2.5)(-2,-0.2)%
\cell{-2}{0.6}{br}{#1}%
\cell{-0.7}{2.2}{br}{#2}%
\cell{0.7}{2.2}{bl}{#3}%
\cell{2}{0.6}{bl}{#4}%
\cell{0}{-0.2}{t}{#5}%
\cell{0}{0.8}{c}{#6}%
\put(-2,0){\line(1,3){0.5}}%
\put(-1.5,1.5){\line(3,2){1.5}}%
\put(0,2.5){\line(3,-2){1.5}}%
\put(1.5,1.5){\line(1,-3){0.5}}%
\put(-2,0){\line(1,0){4}}%
\end{picture}}

\mcm{\topeds}{6}{%
\ginitdims{4}{2.5}%
\sidespic{#1}%
\abovepic{#2}%
\abovepic{#3}%
\sidespic{#4}%
\belowpic{#5}%
\present{\pretopeds{#1}{#2}{#3}{#4}{#5}{#6}}}

\newcommand{\pretopeqs}[5]{%
\begin{picture}(4,2.5)(-2,-0.2)%
\cell{-2}{0.6}{br}{#1}%
\cell{-1}{2.2}{br}{#2}%
\cell{2}{0.6}{bl}{#3}%
\cell{0}{-0.2}{t}{#4}%
\cell{0}{0.8}{c}{#5}%
\put(-2,0){\line(1,3){0.5}}%
\put(-1.5,1.5){\line(1,1){1}}%
\cell{0.9}{2.3}{c}{\ddots}
\put(1.5,1.5){\line(1,-3){0.5}}%
\put(-2,0){\line(1,0){4}}%
\end{picture}}

\mcm{\topeqs}{5}{%
\ginitdims{4}{2.5}%
\sidespic{#1}%
\abovepic{#2}%
\sidespic{#3}%
\belowpic{#4}%
\present{\pretopeqs{#1}{#2}{#3}{#4}{#5}}}

\newcommand{\pretopebases}[1]{%
\begin{picture}(4,0.4)(0,-0.2)%
\cell{2}{0.2}{b}{#1}%
\put(0,0){\line(1,0){4}}%
\end{picture}}

\mcm{\topebases}{1}{%
\ginitdims{4}{0.4}%
\abovepic{#1}%
\present{\pretopebases{#1}}}

\newcommand{\pregdots}[6]{%
\begin{picture}(5,8.4)(0,-2.7)%
\cell{2.5}{5.7}{b}{#1}%
\cell{1.5}{2.8}{b}{#2}%
\cell{1.5}{0.2}{t}{#3}%
\cell{2.5}{-2.7}{t}{#4}%
\cell{2.7}{4.25}{l}{#5}%
\cell{2.7}{-1.25}{l}{#6}%
\qbezier(0,1.5)(2.5,9.5)(5,1.5)%
\qbezier(0,1.5)(2.5,4)(5,1.5)%
\qbezier(0,1.5)(2.5,-1)(5,1.5)%
\qbezier(0,1.5)(2.5,-6.5)(5,1.5)%
\put(2.5,5.25){\vector(0,-1){2}}%
\put(2.5,-0.25){\vector(0,-1){2}}%
\cell{2.5}{1.7}{c}{\vdots}%
\put(5,1.5){\vector(1,-4){0}}%
\put(5,1.5){\vector(4,-3){0}}%
\put(5,1.5){\vector(4,3){0}}%
\put(5,1.5){\vector(1,4){0}}%
\end{picture}}

\mcm{\gdots}{6}{%
\ginitdims{5}{8.4}%
\abovepic{#1}%
\belowpic{#4}%
\present{\pregdots{#1}{#2}{#3}{#4}{#5}{#6}}}

\newlength{\volt}
\makeatletter

\def\diagram{\m@th\leftwidth=\z@ \rightwidth=\z@ \topheight=\z@
\botheight=\z@ \setbox\@picbox\hbox\bgroup}

\def\enddiagram{\egroup\wd\@picbox\rightwidth\unitlength
\ht\@picbox\topheight\unitlength \dp\@picbox\botheight\unitlength
\hskip\leftwidth\unitlength\box\@picbox}

\def\bfig{\begin{diagram}}
\def\efig{\end{diagram}}
\newcount\wideness \newcount\leftwidth \newcount\rightwidth
\newcount\highness \newcount\topheight \newcount\botheight

\def\ratchet#1#2{\ifnum#1<#2 \global #1=#2 \fi}

\def\putbox(#1,#2)#3{%
\horsize{\wideness}{#3} \divide\wideness by 2 {\advance\wideness
by #1 \ratchet{\rightwidth}{\wideness}} {\advance\wideness by -#1
\ratchet{\leftwidth}{\wideness}} \vertsize{\highness}{#3}
\divide\highness by 2 {\advance\highness by #2
\ratchet{\topheight}{\highness}} {\advance\highness by -#2
\ratchet{\botheight}{\highness}} \put(#1,#2){\makebox(0,0){$#3$}}}

\def\putlbox(#1,#2)#3{%
\horsize{\wideness}{#3} {\advance\wideness by #1
\ratchet{\rightwidth}{\wideness}} {\ratchet{\leftwidth}{-#1}}
\vertsize{\highness}{#3} \divide\highness by 2 {\advance\highness
by #2 \ratchet{\topheight}{\highness}} {\advance\highness by -#2
\ratchet{\botheight}{\highness}}
\put(#1,#2){\makebox(0,0)[l]{$#3$}}}

\def\putrbox(#1,#2)#3{%
\horsize{\wideness}{#3} {\ratchet{\rightwidth}{#1}}
{\advance\wideness by -#1 \ratchet{\leftwidth}{\wideness}}
\vertsize{\highness}{#3} \divide\highness by 2 {\advance\highness
by #2 \ratchet{\topheight}{\highness}} {\advance\highness by -#2
\ratchet{\botheight}{\highness}}
\put(#1,#2){\makebox(0,0)[r]{$#3$}}}

\def\adjust[#1]{} 

\newcount \coefa
\newcount \coefb
\newcount \coefc
\newcount\tempcounta
\newcount\tempcountb
\newcount\tempcountc
\newcount\tempcountd
\newcount\xext
\newcount\yext
\newcount\xoff
\newcount\yoff
\newcount\gap%
\newcount\arrowtypea
\newcount\arrowtypeb
\newcount\arrowtypec
\newcount\arrowtyped
\newcount\arrowtypee
\newcount\height
\newcount\width
\newcount\xpos
\newcount\ypos
\newcount\run
\newcount\rise
\newcount\arrowlength
\newcount\halflength
\newcount\arrowtype
\newdimen\tempdimen
\newdimen\xlen
\newdimen\ylen
\newsavebox{\tempboxa}%
\newsavebox{\tempboxb}%
\newsavebox{\tempboxc}%

\newdimen\w@dth

\def\setw@dth#1#2{\setbox\z@\hbox{\m@th$#1$}\w@dth=\wd\z@
\setbox\@ne\hbox{\m@th$#2$}\ifnum\w@dth<\wd\@ne \w@dth=\wd\@ne \fi
\advance\w@dth by 1.2em}

\def\t@^#1_#2{\allowbreak\def\n@one{#1}\def\n@two{#2}\mathrel
{\setw@dth{#1}{#2} \mathop{\hbox to
\w@dth{\rightarrowfill}}\limits \ifx\n@one\empty\else
^{\box\z@}\fi \ifx\n@two\empty\else _{\box\@ne}\fi}}
\def\t@@^#1{\@ifnextchar_{\t@^{#1}}{\t@^{#1}_{}}}
\def\to{\@ifnextchar^{\t@@}{\t@@^{}}}

\def\t@left^#1_#2{\def\n@one{#1}\def\n@two{#2}\mathrel{\setw@dth{#1}{#2}
\mathop{\hbox to \w@dth{\leftarrowfill}}\limits
\ifx\n@one\empty\else ^{\box\z@}\fi \ifx\n@two\empty\else
_{\box\@ne}\fi}}
\def\t@@left^#1{\@ifnextchar_{\t@left^{#1}}{\t@left^{#1}_{}}}
\def\toleft{\@ifnextchar^{\t@@left}{\t@@left^{}}}

\def\two@^#1_#2{\allowbreak
\def\n@one{#1}\def\n@two{#2}\mathrel{\setw@dth{#1}{#2}
\mathop{\vcenter{\lineskip\z@\baselineskip\z@
                 \hbox to \w@dth{\rightarrowfill}%
                 \hbox to \w@dth{\rightarrowfill}}%
       }\limits
\ifx\n@one\empty\else ^{\box\z@}\fi \ifx\n@two\empty\else
_{\box\@ne}\fi}}
\def\tw@@^#1{\@ifnextchar _{\two@^{#1}}{\two@^{#1}_{}}}
\def\two{\@ifnextchar ^{\tw@@}{\tw@@^{}}}

\def\tofr@^#1_#2{\def\n@one{#1}\def\n@two{#2}\mathrel{\setw@dth{#1}{#2}
\mathop{\vcenter{\hbox to \w@dth{\rightarrowfill}\kern-1.7ex
                 \hbox to \w@dth{\leftarrowfill}}%
       }\limits
\ifx\n@one\empty\else ^{\box\z@}\fi \ifx\n@two\empty\else
_{\box\@ne}\fi}}
\def\t@fr@^#1{\@ifnextchar_ {\tofr@^{#1}}{\tofr@^{#1}_{}}}
\def\tofro{\@ifnextchar^ {\t@fr@}{\t@fr@^{}}}

\def\mon{\mathop{\m@th\hbox to
      14.6\P@{\lasyb\char'51\hskip-2.1\P@$\arrext$\hss
$\mathord\rightarrow$}}\limits} 
\def\leftmono{\mathrel{\m@th\hbox to
14.6\P@{$\mathord\leftarrow$\hss$\arrext$\hskip-2.1\P@\lasyb\char'50%
}}\limits} 
\mathchardef\arrext="0200       

\def\settypes(#1,#2,#3){\arrowtypea#1 \arrowtypeb#2 \arrowtypec#3}
\def\settoheight#1#2{\setbox\@tempboxa\hbox{#2}#1\ht\@tempboxa\relax}%
\def\settodepth#1#2{\setbox\@tempboxa\hbox{#2}#1\dp\@tempboxa\relax}%
\def\settokens`#1`#2`#3`#4`{%
     \def\tokena{#1}\def\tokenb{#2}\def\tokenc{#3}\def\tokend{#4}}
\def\setsqparms[#1`#2`#3`#4;#5`#6]{%
\arrowtypea #1 \arrowtypeb #2 \arrowtypec #3 \arrowtyped #4
\width #5 \height #6 }
\def\setpos(#1,#2){\xpos=#1 \ypos#2}

\def\settriparms[#1`#2`#3;#4]{\settripairparms[#1`#2`#3`1`1;#4]}%

\def\settripairparms[#1`#2`#3`#4`#5;#6]{%
\arrowtypea #1 \arrowtypeb #2 \arrowtypec #3 \arrowtyped #4
\arrowtypee #5 \width #6 \height #6 }

\def\resetparms{\settripairparms[1`1`1`1`1;500]\width 500}

\resetparms

\def\mvector(#1,#2)#3{
\put(0,0){\vector(#1,#2){#3}}%
\put(0,0){\vector(#1,#2){26}}%
}
\def\evector(#1,#2)#3{{
\arrowlength #3
\put(0,0){\vector(#1,#2){\arrowlength}}%
\advance \arrowlength by-30
\put(0,0){\vector(#1,#2){\arrowlength}}%
}}

\def\horsize#1#2{%
\settowidth{\tempdimen}{$#2$}%
#1=\tempdimen \divide #1 by\unitlength }

\def\vertsize#1#2{%
\settoheight{\tempdimen}{$#2$}%
#1=\tempdimen
\settodepth{\tempdimen}{$#2$}%
\advance #1 by\tempdimen \divide #1 by\unitlength }

\def\putvector(#1,#2)(#3,#4)#5#6{{%
\ifnum3<\arrowtype \putdashvector(#1,#2)(#3,#4)#5\arrowtype \else
\ifnum\arrowtype<-3 \putdashvector(#1,#2)(#3,#4)#5\arrowtype \else
\xpos=#1 \ypos=#2 \run=#3 \rise=#4 \arrowlength=#5 \ifnum
\arrowtype<0
    \ifnum \run=0
        \advance \ypos by-\arrowlength
    \else
        \tempcounta \arrowlength
        \multiply \tempcounta by\rise
        \divide \tempcounta by\run
        \ifnum\run>0
            \advance \xpos by\arrowlength
            \advance \ypos by\tempcounta
        \else
            \advance \xpos by-\arrowlength
            \advance \ypos by-\tempcounta
        \fi
    \fi
    \multiply \arrowtype by-1
    \multiply \rise by-1
    \multiply \run by-1
\fi \ifcase \arrowtype
\or \put(\xpos,\ypos){\vector(\run,\rise){\arrowlength}}%
\or \put(\xpos,\ypos){\mvector(\run,\rise)\arrowlength}%
\or \put(\xpos,\ypos){\evector(\run,\rise){\arrowlength}}%
\fi\fi\fi }}

\def\putsplitvector(#1,#2)#3#4{
\xpos #1 \ypos #2 \arrowtype #4 \halflength #3 \arrowlength #3
\gap 140 \advance \halflength by-\gap \divide \halflength by2
\ifnum\arrowtype>0
   \ifcase \arrowtype
   \or \put(\xpos,\ypos){\line(0,-1){\halflength}}%
       \advance\ypos by-\halflength
       \advance\ypos by-\gap
       \put(\xpos,\ypos){\vector(0,-1){\halflength}}%
   \or \put(\xpos,\ypos){\line(0,-1)\halflength}%
       \put(\xpos,\ypos){\vector(0,-1)3}%
       \advance\ypos by-\halflength
       \advance\ypos by-\gap
       \put(\xpos,\ypos){\vector(0,-1){\halflength}}%
   \or \put(\xpos,\ypos){\line(0,-1)\halflength}%
       \advance\ypos by-\halflength
       \advance\ypos by-\gap
       \put(\xpos,\ypos){\evector(0,-1){\halflength}}%
   \fi
\else \arrowtype=-\arrowtype
   \ifcase\arrowtype
   \or \advance \ypos by-\arrowlength
       \put(\xpos,\ypos){\line(0,1){\halflength}}%
       \advance\ypos by\halflength
       \advance\ypos by\gap
       \put(\xpos,\ypos){\vector(0,1){\halflength}}%
   \or \advance \ypos by-\arrowlength
       \put(\xpos,\ypos){\line(0,1)\halflength}%
       \put(\xpos,\ypos){\vector(0,1)3}%
       \advance\ypos by\halflength
       \advance\ypos by\gap
       \put(\xpos,\ypos){\vector(0,1){\halflength}}%
   \or \advance \ypos by-\arrowlength
       \put(\xpos,\ypos){\line(0,1)\halflength}%
       \advance\ypos by\halflength
       \advance\ypos by\gap
       \put(\xpos,\ypos){\evector(0,1){\halflength}}%
   \fi
\fi }

\def\putmorphism(#1)(#2,#3)[#4`#5`#6]#7#8#9{{%
\run #2 \rise #3 \ifnum\rise=0
  \puthmorphism(#1)[#4`#5`#6]{#7}{#8}#9%
\else\ifnum\run=0
  \putvmorphism(#1)[#4`#5`#6]{#7}{#8}#9%
\else
\setpos(#1)%
\arrowlength #7 \arrowtype #8 \ifnum\run=0 \else\ifnum\rise=0
\else \ifnum\run>0
    \coefa=1
\else
   \coefa=-1
\fi \ifnum\arrowtype>0
   \coefb=0
   \coefc=-1
\else
   \coefb=\coefa
   \coefc=1
   \arrowtype=-\arrowtype
\fi \width=2 \multiply \width by\run \divide \width by\rise
\ifnum \width<0  \width=-\width\fi \advance\width by60 \if l#9
\width=-\width\fi
\putbox(\xpos,\ypos){#4}
{\multiply \coefa by\arrowlength
\advance\xpos by\coefa \multiply \coefa by\rise \divide \coefa
by\run \advance \ypos by\coefa
\putbox(\xpos,\ypos){#5} }%
{\multiply \coefa by\arrowlength
\divide \coefa by2 \advance \xpos by\coefa \advance \xpos by\width
\multiply \coefa by\rise \divide \coefa by\run \advance \ypos
by\coefa
\if l#9%
   \putrbox(\xpos,\ypos){#6}%
\else\if r#9%
   \putlbox(\xpos,\ypos){#6}%
\fi\fi }%
{\multiply \rise by-\coefc
\multiply \run by-\coefc \multiply \coefb by\arrowlength \advance
\xpos by\coefb \multiply \coefb by\rise \divide \coefb by\run
\advance \ypos by\coefb \multiply \coefc by70 \advance \ypos
by\coefc \multiply \coefc by\run \divide \coefc by\rise \advance
\xpos by\coefc \multiply \coefa by140 \multiply \coefa by\run
\divide \coefa by\rise \advance \arrowlength by\coefa
\ifcase\arrowtype
\or \put(\xpos,\ypos){\vector(\run,\rise){\arrowlength}}%
\or \put(\xpos,\ypos){\mvector(\run,\rise){\arrowlength}}%
\or \put(\xpos,\ypos){\evector(\run,\rise){\arrowlength}}%
\fi}\fi\fi\fi\fi}}

\newcount\numbdashes \newcount\lengthdash \newcount\increment

\def\howmanydashes{
\numbdashes=\arrowlength \lengthdash=40 \divide\numbdashes by
\lengthdash \lengthdash=\arrowlength \divide\lengthdash by
\numbdashes
\increment=\lengthdash \multiply\lengthdash by 3
\divide\lengthdash by 5 }

\def\putdashvector(#1)(#2,#3)#4#5{%
\ifnum#3=0 \putdashhvector(#1){#4}#5 \else \ifnum#2=0
\putdashvvector(#1){#4}#5\fi\fi}

\def\putdashhvector(#1,#2)#3#4{{%
\arrowlength=#3 \howmanydashes
\multiput(#1,#2)(\increment,0){\numbdashes}%
{\vrule height .4pt width \lengthdash\unitlength} \arrowtype=#4
\xpos=#1 \ifnum\arrowtype<0 \advance\arrowtype by 7 \fi
\ifcase\arrowtype \or \advance\xpos by 10
    \put(\xpos,#2){\vector(-1,0){\lengthdash}}
    \advance\xpos by 40
    \put(\xpos,#2){\vector(-1,0){\lengthdash}}
\or \advance \xpos by 10
    \put(\xpos,#2){\vector(-1,0){\lengthdash}}
    \advance\xpos by  \arrowlength
    \advance\xpos by  -50
    \put(\xpos,#2){\vector(-1,0){\lengthdash}}
\or \advance\xpos by 10
    \put(\xpos,#2){\vector(-1,0){\lengthdash}}
\or \advance\xpos by \arrowlength
    \advance\xpos by -\lengthdash
    \put(\xpos,#2){\vector(1,0){\lengthdash}}
\or {\advance\xpos by 10
    \put(\xpos,#2){\vector(1,0){\lengthdash}}}
    \advance\xpos by \arrowlength
    \advance\xpos by -\lengthdash
    \put(\xpos,#2){\vector(1,0){\lengthdash}}
\or \advance\xpos by \arrowlength
    \advance\xpos by -\lengthdash
    \put(\xpos,#2){\vector(1,0){\lengthdash}}
    \advance\xpos by -40
    \put(\xpos,#2){\vector(1,0){\lengthdash}}
   \fi
}}

\def\putdashvvector(#1,#2)#3#4{{%
\arrowlength=#3 \howmanydashes \ypos=#2 \advance\ypos by
-\arrowlength
\multiput(#1,#2)(0,\increment){\numbdashes}%
    {\vrule width .4pt height \lengthdash\unitlength}
\arrowtype=#4 \ypos=#2 \ifnum\arrowtype<0 \advance\arrowtype by 7
\fi \ifcase\arrowtype \or \advance\ypos by \arrowlength
\advance\ypos by -40
    \put(#1,\ypos){\vector(0,1){\lengthdash}}
    \advance\ypos by -40
    \put(#1,\ypos){\vector(0,1){\lengthdash}}
\or \advance\ypos by 10
    \put(#1,\ypos){\vector(0,1){\lengthdash}}
    \advance\ypos by \arrowlength \advance\ypos by -40
    \put(#1,\ypos){\vector(0,1){\lengthdash}}
\or \advance\ypos by \arrowlength \advance\ypos by -40
    \put(#1,\ypos){\vector(0,1){\lengthdash}}
\or \advance\ypos by 10
    \put(#1,\ypos){\vector(0,-1){\lengthdash}}
\or \advance\ypos by 10
    \put(#1,\ypos){\vector(0,-1){\lengthdash}}
    \advance\ypos by \arrowlength \advance\ypos by -40
    \put(#1,\ypos){\vector(0,-1){\lengthdash}}
\or \advance\ypos by 10
    \put(#1,\ypos){\vector(0,-1){\lengthdash}}
    \advance\ypos by 40
    \put(#1,\ypos){\vector(0,-1){\lengthdash}}
\fi }}

\def\puthmorphism(#1,#2)[#3`#4`#5]#6#7#8{{%
\xpos #1 \ypos #2 \width #6 \arrowlength #6 \arrowtype=#7
\putbox(\xpos,\ypos){#3\vphantom{#4}}%
{\advance \xpos by\arrowlength
\putbox(\xpos,\ypos){\vphantom{#3}#4}}%
\horsize{\tempcounta}{#3}%
\horsize{\tempcountb}{#4}%
\divide \tempcounta by2 \divide \tempcountb by2 \advance
\tempcounta by30 \advance \tempcountb by30 \advance \xpos
by\tempcounta \advance \arrowlength by-\tempcounta \advance
\arrowlength by-\tempcountb
\putvector(\xpos,\ypos)(1,0)\arrowlength\arrowtype \divide
\arrowlength by2 \advance \xpos by\arrowlength
\vertsize{\tempcounta}{#5}%
\divide\tempcounta by2 \advance \tempcounta by20
\if a#8 %
   \advance \ypos by\tempcounta
   \putbox(\xpos,\ypos){#5}%
\else
   \advance \ypos by-\tempcounta
   \putbox(\xpos,\ypos){#5}%
\fi}}

\def\putvmorphism(#1,#2)[#3`#4`#5]#6#7#8{{%
\xpos #1 \ypos #2 \arrowlength #6 \arrowtype #7
\settowidth{\xlen}{$#5$}%
\putbox(\xpos,\ypos){#3}%
{\advance \ypos by-\arrowlength
\putbox(\xpos,\ypos){#4}}%
{\advance\arrowlength by-140 \advance \ypos by-70 \ifdim\xlen>0pt
   \if m#8%
      \putsplitvector(\xpos,\ypos)\arrowlength\arrowtype
   \else
   \putvector(\xpos,\ypos)(0,-1)\arrowlength\arrowtype
   \fi
\else
   \putvector(\xpos,\ypos)(0,-1)\arrowlength\arrowtype
\fi}%
\ifdim\xlen>0pt
   \divide \arrowlength by2
   \advance\ypos by-\arrowlength
   \if l#8%
      \advance \xpos by-40
      \putrbox(\xpos,\ypos){#5}%
   \else\if r#8%
      \advance \xpos by40
      \putlbox(\xpos,\ypos){#5}%
   \else
      \putbox(\xpos,\ypos){#5}%
   \fi\fi
\fi }}

\def\putsquarep<#1>(#2)[#3;#4`#5`#6`#7]{{%
\setsqparms[#1]%
\setpos(#2)%
\settokens`#3`%
\puthmorphism(\xpos,\ypos)[\tokenc`\tokend`{#7}]{\width}{\arrowtyped}b%
\advance\ypos by \height
\puthmorphism(\xpos,\ypos)[\tokena`\tokenb`{#4}]{\width}{\arrowtypea}a%
\putvmorphism(\xpos,\ypos)[``{#5}]{\height}{\arrowtypeb}l%
\advance\xpos by \width
\putvmorphism(\xpos,\ypos)[``{#6}]{\height}{\arrowtypec}r%
}}

\def\putsquare{\@ifnextchar <{\putsquarep}{\putsquarep%
   <\arrowtypea`\arrowtypeb`\arrowtypec`\arrowtyped;\width`\height>}}
\def\square{\@ifnextchar< {\squarep}{\squarep
   <\arrowtypea`\arrowtypeb`\arrowtypec`\arrowtyped;\width`\height>}}
\def\squarep<#1>[#2`#3`#4`#5;#6`#7`#8`#9]{{
\setsqparms[#1]
\diagram
\putsquarep<\arrowtypea`\arrowtypeb`\arrowtypec`
\arrowtyped;\width`\height>
(0,0)[#2`#3`#4`{#5};#6`#7`#8`{#9}]
\enddiagram
}}                                                 
\def\putptrianglep<#1>(#2,#3)[#4`#5`#6;#7`#8`#9]{{%
\settriparms[#1]%
\xpos=#2 \ypos=#3 \advance\ypos by \height
\puthmorphism(\xpos,\ypos)[#4`#5`{#7}]{\height}{\arrowtypea}a%
\putvmorphism(\xpos,\ypos)[`#6`{#8}]{\height}{\arrowtypeb}l%
\advance\xpos by\height
\putmorphism(\xpos,\ypos)(-1,-1)[``{#9}]{\height}{\arrowtypec}r%
}}

\def\putptriangle{\@ifnextchar <{\putptrianglep}{\putptrianglep
   <\arrowtypea`\arrowtypeb`\arrowtypec;\height>}}
\def\ptriangle{\@ifnextchar <{\ptrianglep}{\ptrianglep
   <\arrowtypea`\arrowtypeb`\arrowtypec;\height>}}
\def\ptrianglep<#1>[#2`#3`#4;#5`#6`#7]{{
\settriparms[#1]
\diagram
\putptrianglep<\arrowtypea`\arrowtypeb`
\arrowtypec;\height>
(0,0)[#2`#3`#4;#5`#6`{#7}]
\enddiagram
}}                                            

\def\putqtrianglep<#1>(#2,#3)[#4`#5`#6;#7`#8`#9]{{%
\settriparms[#1]%
\xpos=#2 \ypos=#3 \advance\ypos by\height
\puthmorphism(\xpos,\ypos)[#4`#5`{#7}]{\height}{\arrowtypea}a%
\putmorphism(\xpos,\ypos)(1,-1)[``{#8}]{\height}{\arrowtypeb}l%
\advance\xpos by\height
\putvmorphism(\xpos,\ypos)[`#6`{#9}]{\height}{\arrowtypec}r%
}}

\def\putqtriangle{\@ifnextchar <{\putqtrianglep}{\putqtrianglep
   <\arrowtypea`\arrowtypeb`\arrowtypec;\height>}}
\def\qtriangle{\@ifnextchar <{\qtrianglep}{\qtrianglep
   <\arrowtypea`\arrowtypeb`\arrowtypec;\height>}}
\def\qtrianglep<#1>[#2`#3`#4;#5`#6`#7]{{
\settriparms[#1]
\width=\height                                
\diagram
\putqtrianglep<\arrowtypea`\arrowtypeb`
\arrowtypec;\height>
(0,0)[#2`#3`#4;#5`#6`{#7}]
\enddiagram
}}

\def\putdtrianglep<#1>(#2,#3)[#4`#5`#6;#7`#8`#9]{{%
\settriparms[#1]%
\xpos=#2 \ypos=#3
\puthmorphism(\xpos,\ypos)[#5`#6`{#9}]{\height}{\arrowtypec}b%
\advance\xpos by \height \advance\ypos by\height
\putmorphism(\xpos,\ypos)(-1,-1)[``{#7}]{\height}{\arrowtypea}l%
\putvmorphism(\xpos,\ypos)[#4``{#8}]{\height}{\arrowtypeb}r%
}}

\def\putdtriangle{\@ifnextchar <{\putdtrianglep}{\putdtrianglep
   <\arrowtypea`\arrowtypeb`\arrowtypec;\height>}}
\def\dtriangle{\@ifnextchar <{\dtrianglep}{\dtrianglep
   <\arrowtypea`\arrowtypeb`\arrowtypec;\height>}}
\def\dtrianglep<#1>[#2`#3`#4;#5`#6`#7]{{
\settriparms[#1]
\width=\height                                
\diagram
\putdtrianglep<\arrowtypea`\arrowtypeb`
\arrowtypec;\height>
(0,0)[#2`#3`#4;#5`#6`{#7}]
\enddiagram
}}

\def\putbtrianglep<#1>(#2,#3)[#4`#5`#6;#7`#8`#9]{{%
\settriparms[#1]%
\xpos=#2 \ypos=#3
\puthmorphism(\xpos,\ypos)[#5`#6`{#9}]{\height}{\arrowtypec}b%
\advance\ypos by\height
\putmorphism(\xpos,\ypos)(1,-1)[``{#8}]{\height}{\arrowtypeb}r%
\putvmorphism(\xpos,\ypos)[#4``{#7}]{\height}{\arrowtypea}l%
}}

\def\putbtriangle{\@ifnextchar <{\putbtrianglep}{\putbtrianglep
   <\arrowtypea`\arrowtypeb`\arrowtypec;\height>}}
\def\btriangle{\@ifnextchar <{\btrianglep}{\btrianglep
   <\arrowtypea`\arrowtypeb`\arrowtypec;\height>}}
\def\btrianglep<#1>[#2`#3`#4;#5`#6`#7]{{
\settriparms[#1]
\width=\height                               
\diagram
\putbtrianglep<\arrowtypea`\arrowtypeb`
\arrowtypec;\height>
(0,0)[#2`#3`#4;#5`#6`{#7}]
\enddiagram
}}

\def\putAtrianglep<#1>(#2,#3)[#4`#5`#6;#7`#8`#9]{{%
\settriparms[#1]%
\xpos=#2 \ypos=#3 {\multiply \height by2
\puthmorphism(\xpos,\ypos)[#5`#6`{#9}]{\height}{\arrowtypec}b}%
\advance\xpos by\height \advance\ypos by\height
\putmorphism(\xpos,\ypos)(-1,-1)[#4``{#7}]{\height}{\arrowtypea}l%
\putmorphism(\xpos,\ypos)(1,-1)[``{#8}]{\height}{\arrowtypeb}r%
}}

\def\putAtriangle{\@ifnextchar <{\putAtrianglep}{\putAtrianglep
   <\arrowtypea`\arrowtypeb`\arrowtypec;\height>}}
\def\Atriangle{\@ifnextchar <{\Atrianglep}{\Atrianglep
   <\arrowtypea`\arrowtypeb`\arrowtypec;\height>}}
\def\Atrianglep<#1>[#2`#3`#4;#5`#6`#7]{{
\settriparms[#1]
\width=\height                                     
\diagram
\putAtrianglep<\arrowtypea`\arrowtypeb`
\arrowtypec;\height>
(0,0)[#2`#3`#4;#5`#6`{#7}]
\enddiagram
}}

\def\putAtrianglepairp<#1>(#2)[#3;#4`#5`#6`#7`#8]{{%
\settripairparms[#1]%
\setpos(#2)%
\settokens`#3`%
\puthmorphism(\xpos,\ypos)[\tokenb`\tokenc`{#7}]{\height}{\arrowtyped}b%
\advance\xpos by\height
\puthmorphism(\xpos,\ypos)[\phantom{\tokenc}`\tokend`{#8}]%
{\height}{\arrowtypee}b%
\advance\ypos by\height
\putmorphism(\xpos,\ypos)(-1,-1)[\tokena``{#4}]{\height}{\arrowtypea}l%
\putvmorphism(\xpos,\ypos)[``{#5}]{\height}{\arrowtypeb}m%
\putmorphism(\xpos,\ypos)(1,-1)[``{#6}]{\height}{\arrowtypec}r%
}}

\def\putAtrianglepair{\@ifnextchar <{\putAtrianglepairp}{\putAtrianglepairp%
   <\arrowtypea`\arrowtypeb`\arrowtypec`\arrowtyped`\arrowtypee;\height>}}
\def\Atrianglepair{\@ifnextchar <{\Atrianglepairp}{\Atrianglepairp%
   <\arrowtypea`\arrowtypeb`\arrowtypec`\arrowtyped`\arrowtypee;\height>}}

\def\Atrianglepairp<#1>[#2;#3`#4`#5`#6`#7]{{
\settripairparms[#1]
\settokens`#2`
\width=\height                                
\diagram
\putAtrianglepairp                            
<\arrowtypea`\arrowtypeb`\arrowtypec`
\arrowtyped`\arrowtypee;\height>
(0,0)[{#2};#3`#4`#5`#6`{#7}]
\enddiagram
}}

\def\putVtrianglep<#1>(#2,#3)[#4`#5`#6;#7`#8`#9]{{%
\settriparms[#1]%
\xpos=#2 \ypos=#3 \advance\ypos by\height {\multiply\height by2
\puthmorphism(\xpos,\ypos)[#4`#5`{#7}]{\height}{\arrowtypea}a}%
\putmorphism(\xpos,\ypos)(1,-1)[`#6`{#8}]{\height}{\arrowtypeb}l%
\advance\xpos by\height \advance\xpos by\height
\putmorphism(\xpos,\ypos)(-1,-1)[``{#9}]{\height}{\arrowtypec}r%
}}

\def\putVtriangle{\@ifnextchar <{\putVtrianglep}{\putVtrianglep
   <\arrowtypea`\arrowtypeb`\arrowtypec;\height>}}
\def\Vtriangle{\@ifnextchar <{\Vtrianglep}{\Vtrianglep
   <\arrowtypea`\arrowtypeb`\arrowtypec;\height>}}
\def\Vtrianglep<#1>[#2`#3`#4;#5`#6`#7]{{
\settriparms[#1]
\width=\height                                 
\diagram
\putVtrianglep<\arrowtypea`\arrowtypeb`
\arrowtypec;\height>
(0,0)[#2`#3`#4;#5`#6`{#7}]
\enddiagram
}}

\def\putVtrianglepairp<#1>(#2)[#3;#4`#5`#6`#7`#8]{{
\settripairparms[#1]%
\setpos(#2)%
\settokens`#3`%
\advance\ypos by\height
\putmorphism(\xpos,\ypos)(1,-1)[`\tokend`{#6}]{\height}{\arrowtypec}l%
\puthmorphism(\xpos,\ypos)[\tokena`\tokenb`{#4}]{\height}{\arrowtypea}a%
\advance\xpos by\height
\puthmorphism(\xpos,\ypos)[\phantom{\tokenb}`\tokenc`{#5}]%
{\height}{\arrowtypeb}a%
\putvmorphism(\xpos,\ypos)[``{#7}]{\height}{\arrowtyped}m%
\advance\xpos by\height
\putmorphism(\xpos,\ypos)(-1,-1)[``{#8}]{\height}{\arrowtypee}r%
}}

\def\putVtrianglepair{\@ifnextchar <{\putVtrianglepairp}{\putVtrianglepairp%
    <\arrowtypea`\arrowtypeb`\arrowtypec`\arrowtyped`\arrowtypee;\height>}}
\def\Vtrianglepair{\@ifnextchar <{\Vtrianglepairp}{\Vtrianglepairp%
    <\arrowtypea`\arrowtypeb`\arrowtypec`\arrowtyped`\arrowtypee;\height>}}
\def\Vtrianglepairp<#1>[#2;#3`#4`#5`#6`#7]{{
\settripairparms[#1]
\settokens`#2`
\diagram
\putVtrianglepairp                             
<\arrowtypea`\arrowtypeb`\arrowtypec`
\arrowtyped`\arrowtypee;\height>
(0,0)[{#2};#3`#4`#5`#6`{#7}]
\enddiagram
}}

\def\putCtrianglep<#1>(#2,#3)[#4`#5`#6;#7`#8`#9]{{%
\settriparms[#1]%
\xpos=#2 \ypos=#3 \advance\ypos by\height
\putmorphism(\xpos,\ypos)(1,-1)[``{#9}]{\height}{\arrowtypec}l%
\advance\xpos by\height \advance\ypos by\height
\putmorphism(\xpos,\ypos)(-1,-1)[#4`#5`{#7}]{\height}{\arrowtypea}l%
{\multiply\height by 2
\putvmorphism(\xpos,\ypos)[`#6`{#8}]{\height}{\arrowtypeb}r}%
}}

\def\putCtriangle{\@ifnextchar <{\putCtrianglep}{\putCtrianglep
    <\arrowtypea`\arrowtypeb`\arrowtypec;\height>}}
\def\Ctriangle{\@ifnextchar <{\Ctrianglep}{\Ctrianglep
    <\arrowtypea`\arrowtypeb`\arrowtypec;\height>}}
\def\Ctrianglep<#1>[#2`#3`#4;#5`#6`#7]{{
\settriparms[#1]
\width=\height                               
\diagram
\putCtrianglep<\arrowtypea`\arrowtypeb`
\arrowtypec;\height>
(0,0)[#2`#3`#4;#5`#6`{#7}]
\enddiagram
}}                                           
\def\putDtrianglep<#1>(#2,#3)[#4`#5`#6;#7`#8`#9]{{%
\settriparms[#1]%
\xpos=#2 \ypos=#3 \advance\xpos by\height \advance\ypos by\height
\putmorphism(\xpos,\ypos)(-1,-1)[``{#9}]{\height}{\arrowtypec}r%
\advance\xpos by-\height \advance\ypos by\height
\putmorphism(\xpos,\ypos)(1,-1)[`#5`{#8}]{\height}{\arrowtypeb}r%
{\multiply\height by 2
\putvmorphism(\xpos,\ypos)[#4`#6`{#7}]{\height}{\arrowtypea}l}%
}}

\def\putDtriangle{\@ifnextchar <{\putDtrianglep}{\putDtrianglep
    <\arrowtypea`\arrowtypeb`\arrowtypec;\height>}}
\def\Dtriangle{\@ifnextchar <{\Dtrianglep}{\Dtrianglep
   <\arrowtypea`\arrowtypeb`\arrowtypec;\height>}}
\def\Dtrianglep<#1>[#2`#3`#4;#5`#6`#7]{{
\settriparms[#1]
\width=\height                              
\diagram
\putDtrianglep<\arrowtypea`\arrowtypeb`
\arrowtypec;\height>
(0,0)[#2`#3`#4;#5`#6`{#7}]
\enddiagram
}}                                          
\def\setrecparms[#1`#2]{\width=#1 \height=#2}%

\def\recursep<#1`#2>[#3;#4`#5`#6`#7`#8]{{\m@th
\width=#1 \height=#2 \settokens`#3`
\settowidth{\tempdimen}{$\tokena$} \ifdim\tempdimen=0pt
  \savebox{\tempboxa}{\hbox{$\tokenb$}}%
  \savebox{\tempboxb}{\hbox{$\tokend$}}%
  \savebox{\tempboxc}{\hbox{$#6$}}%
\else
  \savebox{\tempboxa}{\hbox{$\hbox{$\tokena$}\times\hbox{$\tokenb$}$}}%
  \savebox{\tempboxb}{\hbox{$\hbox{$\tokena$}\times\hbox{$\tokend$}$}}%
  \savebox{\tempboxc}{\hbox{$\hbox{$\tokena$}\times\hbox{$#6$}$}}%
\fi \ypos=\height \divide\ypos by 2 \xpos=\ypos \advance\xpos by
\width \bfig
\putCtrianglep<-1`1`1;\ypos>(0,0)[`\tokenc`;#5`#6`{#7}]%
\puthmorphism(\ypos,0)[\tokend`\usebox{\tempboxb}`{#8}]{\width}{-1}b%
\puthmorphism(\ypos,\height)[\tokenb`\usebox{\tempboxa}`{#4}]{\width}{-1}a%
\advance\ypos by \width
\putvmorphism(\ypos,\height)[``\usebox{\tempboxc}]{\height}1r%
\efig }}

\def\recurse{\@ifnextchar <{\recursep}{\recursep<\width`\height>}}

\def\puttwohmorphisms(#1,#2)[#3`#4;#5`#6]#7#8#9{{%
\puthmorphism(#1,#2)[#3`#4`]{#7}0a \ypos=#2 \advance\ypos by 20
\puthmorphism(#1,\ypos)[\phantom{#3}`\phantom{#4}`#5]{#7}{#8}a
\advance\ypos by -40
\puthmorphism(#1,\ypos)[\phantom{#3}`\phantom{#4}`#6]{#7}{#9}b }}

\def\puttwovmorphisms(#1,#2)[#3`#4;#5`#6]#7#8#9{{%
\putvmorphism(#1,#2)[#3`#4`]{#7}0a \xpos=#1 \advance\xpos by -20
\putvmorphism(\xpos,#2)[\phantom{#3}`\phantom{#4}`#5]{#7}{#8}l
\advance\xpos by 40
\putvmorphism(\xpos,#2)[\phantom{#3}`\phantom{#4}`#6]{#7}{#9}r }}

\def\puthcoequalizer(#1)[#2`#3`#4;#5`#6`#7]#8#9{{%
\setpos(#1)%
\puttwohmorphisms(\xpos,\ypos)[#2`#3;#5`#6]{#8}11%
\advance\xpos by #8
\puthmorphism(\xpos,\ypos)[\phantom{#3}`#4`#7]{#8}1{#9} }}

\def\putvcoequalizer(#1)[#2`#3`#4;#5`#6`#7]#8#9{{%
\setpos(#1)%
\puttwovmorphisms(\xpos,\ypos)[#2`#3;#5`#6]{#8}11%
\advance\ypos by -#8
\putvmorphism(\xpos,\ypos)[\phantom{#3}`#4`#7]{#8}1{#9} }}

\def\putthreehmorphisms(#1)[#2`#3;#4`#5`#6]#7(#8)#9{{%
\setpos(#1) \settypes(#8)
\if a#9 %
     \vertsize{\tempcounta}{#5}%
     \vertsize{\tempcountb}{#6}%
     \ifnum \tempcounta<\tempcountb \tempcounta=\tempcountb \fi
\else
     \vertsize{\tempcounta}{#4}%
     \vertsize{\tempcountb}{#5}%
     \ifnum \tempcounta<\tempcountb \tempcounta=\tempcountb \fi
\fi \advance \tempcounta by 60
\puthmorphism(\xpos,\ypos)[#2`#3`#5]{#7}{\arrowtypeb}{#9}
\advance\ypos by \tempcounta
\puthmorphism(\xpos,\ypos)[\phantom{#2}`\phantom{#3}`#4]{#7}{\arrowtypea}{#9}
\advance\ypos by -\tempcounta \advance\ypos by -\tempcounta
\puthmorphism(\xpos,\ypos)[\phantom{#2}`\phantom{#3}`#6]{#7}{\arrowtypec}{#9}
}}

\def\setarrowtoks[#1`#2`#3`#4`#5`#6]{%
\def\toka{#1}
\def\tokb{#2}
\def\tokc{#3}
\def\tokd{#4}
\def\toke{#5}
\def\tokf{#6}
}
\def\hex{\@ifnextchar <{\hexp}{\hexp<1000`400>}}
\def\hexp<#1`#2>[#3`#4`#5`#6`#7`#8;#9]{%
\setarrowtoks[#9] \yext=#2 \advance \yext by #2 \xext=#1
\advance\xext by \yext \bfig
\putCtriangle<-1`0`1;#2>(0,0)[`#5`;\tokb``\tokd] \xext=#1
\yext=#2 \advance \yext by #2
\putsquare<1`0`0`1;\xext`\yext>(#2,0)[#3`#4`#7`#8;\toka```\tokf]
\advance \xext by #2
\putDtriangle<0`1`-1;#2>(\xext,0)[`#6`;`\tokc`\toke] \efig }

\makeatother

\input{tcilatex}

\newcommand{\cL}{{\cal L}}
\newcommand{\ve}{\varepsilon}
\newcommand{\bth}{{\Theta}}
\newcommand{\bla}{{\Lambda}}
\newcommand{\wt}{\widetilde}
\newcommand{\ot}{\otimes}
\newcommand{\om}{\omega}
\newcommand{\Y}{Y\to X}
\newcommand{\la}{\lambda}
\newcommand{\al}{\alpha}
\newcommand{\bt}{\beta}
\newcommand{\w}{\wedge}
\newcommand{\cO}{{\bf\Omega}}
\newcommand{\m}{\mu}
\newcommand{\dr}{\partial}
\newcommand{\bL}{{\frak L}}
\newcommand{\cH}{{\cal H}}
\newcommand{\wh}{\widehat}
\newcommand{\g}{\gamma}
\newcommand{\G}{\Gamma}
\newcommand{\beq}{\begin{equation}}
\newcommand{\eeq}{\end{equation}}
\newcommand{\be}{\begin{eqnarray*}}
\newcommand{\ee}{\end{eqnarray*}}

\def\op#1{\mathop{\fam0 #1}\limits}

\begin{document}

\title{Time\,\&\,Fitness--Dependent Hamiltonian Biomechanics}\author{Tijana T. Ivancevic\\ {\small Society for Nonlinear Dynamics in Human Factors, Adelaide, Australia}\\
{\small and}\\
{\small CITECH Research IP Pty Ltd, Adelaide, Australia}\\
{\small e-mail: ~tijana.ivancevic@alumni.adelaide.edu.au}}\date{}\maketitle

\tableofcontents

\begin{abstract}
In this paper we propose the time\,\&\,fitness-dependent Hamiltonian form of human biomechanics, in which \emph{total mechanical + biochemical energy is not conserved}. Starting with the Covariant Force Law, we first develop autonomous Hamiltonian biomechanics. Then we extend it using a powerful geometrical machinery consisting of fibre bundles, jet manifolds, polysymplectic geometry and Hamiltonian connections. In this way we derive time-dependent dissipative Hamiltonian equations and the fitness evolution equation for the general time\,\&\,fitness-dependent human biomechanical system.\\

\end{abstract}

\section{Introduction}

Most of dynamics in both classical and quantum physics is based on \emph{assumption of a total energy conservation} (see, e.g. \cite{GaneshADG}). Dynamics based on this assumption of time-independent energy, usually given by Hamiltonian (or Lagrangian) energy function, is called \emph{autonomous}. This basic assumption is naturally inherited in human biomechanics, formally developed using Newton--Euler, Lagrangian or Hamiltonian formalisms (see \cite{GaneshSprSml,GaneshWSc,GaneshSprBig,StrAttr,TijIJHR,TijNis,TijNL,TijSpr}).

And this works fine for most individual movement simulations and predictions, in which the total human energy dissipations are insignificant. However, if we analyze a 100\,m-dash sprinting motion, which is in case of top athletes finished under 10\,s, we can recognize a significant slow-down after about 70\,m in \emph{all} athletes -- despite of their strong intention to finish and win the race, which is an obvious sign of the total energy dissipation. This can be
seen, for example, in a current record-braking speed–distance curve of Usain Bolt, the  world-record holder with 9.69\;s, or in a former record-braking speed–distance curve of Carl Lewis, the former world-record holder (and 9 time Olympic gold medalist) with 9.86\;s (see Figure 3.7 in \cite{TijSpr}). In other words, the \emph{total mechanical + biochemical energy} of a sprinter \emph{cannot be conserved} even for 10\,s. So, if we want to develop a realistic model of intensive human motion that is longer than 7--8\,s (not to speak for instance of a 4 hour tennis match), we necessarily need to use the more advanced formalism of time-dependent mechanics.

In this paper, we will first develop the autonomous Hamiltonian biomechanics as a Hamiltonian representation of the covariant force law [see (\ref{ivcov}) in the next section] and the corresponding covariant force functor. After that we will extend the autonomous Hamiltonian biomechanics into the time-dependent one, in which \emph{total mechanical + biochemical energy is not conserved}. for this, we will use the modern geometric formalism of jet manifolds and bundles.

\section{The Covariant Force Law}

Autonomous Hamiltonian biomechanics (as well as autonomous Lagrangian biomechanics),
based on the postulate of conservation of the total mechanical energy, can be
derived from the \textit{covariant
force law} \cite{GaneshSprSml,GaneshWSc,GaneshSprBig,GaneshADG},
which in `plain English' states:
$$\text{Force 1--form}=\text{Mass distribution}\times \text{Acceleration vector-field},$$
and formally reads (using Einstein's summation convention over repeated indices):
\begin{equation}
F_{i}=m_{ij}a^{j}.  \label{ivcov}
\end{equation}
Here, the force 1--form $F_{i}=F_{i}(t,q,p)=F'_{i}(t,q,\dot{q}),~(i=1,...,n)$ denotes any type of torques and forces acting on a human skeleton,
including excitation and contraction dynamics of
muscular--actuators \cite{Hill,Wilkie,Hatze} and rotational dynamics of hybrid robot
actuators, as well as (nonlinear) dissipative joint torques and
forces and external stochastic perturbation torques and forces \cite{StrAttr}.
$m_{ij}$ is the material (mass--inertia) metric tensor, which gives the total mass distribution of the human body, by including all segmental masses and their individual inertia tensors. $a^{j}$ is the total acceleration vector-field, including all segmental vector-fields, defined as the absolute (Bianchi) derivative $\dot{\bar v}^i$ of all the segmental angular and linear velocities $v^i=\dot{x}^i,~(i=1,...,n)$, where $n$ is the total number of active degrees of freedom (DOF) with local coordinates $(x^i)$.

More formally, this \emph{central Law of biomechanics} represents the \textit{covariant force functor} $\mathcal{F}_*$ defined by the commutative diagram:

\begin{equation}{\large
\putsquare<1`-1`-1`0;1100`500>(360,500)[TT^*M`TTM``
;\mathcal{F}_*`F_i= \dot{p}_i`a^i= \dot{\bar{v}}^i`]
\Vtriangle<0`-1`-1;>[T^*M= \{x^i,p_i\}\;\;`\;\;TM=
\{x^i,v^i\}`\;\;M= \{x^i\};`p_i `v^i= \dot{x}^i] } \label{covfun}
\end{equation}\smallskip

Here, $M\equiv M^n=\{x^i,~(i=1,...,n)\}$ is the biomechanical configuration $n-$manifold, that is the set of all active DOF of the biomechanical system under consideration (in general, human skeleton), with local coordinates $(x^i)$.

The right-hand branch of the fundamental covariant force functor $\mathcal{F}_*:TT^*M \to TTM$ depicted in (\ref{covfun}) is Lagrangian dynamics with its Riemannian geometry. To each $n-$dimensional ($n$D) smooth
manifold $M$ there is associated its $2n$D {\it velocity phase-space
manifold}, denoted by $TM$ and called the tangent bundle of $M$.
The original configuration manifold $M$ is called the {\it base} of $TM$.
There is an onto map $\pi :TM\rightarrow M$, called the
{\it projection}. Above each point $x\in M$ there is a {tangent space}
$T_{x}M=\pi ^{-1}(x)$ to $M$ at $x$, which
is called a {fibre}. The fibre $T_{x}M\subset TM$ is the subset of $%
TM $, such that the total tangent bundle,
$TM=\dbigsqcup\limits_{m\in M}T_{x}M$, is a {disjoint union} of
tangent spaces $T_{x}M$ to $M$ for all points $x\in M$. From
dynamical perspective, the most important quantity in the tangent
bundle concept is the smooth map $v:M\rightarrow TM$, which
is an inverse to the projection $\pi $, i.e, $\pi \circ v=\func{Id}%
_{M},\;\pi (v(x))=x$. It is called the {\it velocity vector-field} $v^i= \dot{x}^i$.\footnote{This explains the dynamical term {\it velocity phase--space}, given to the tangent bundle $TM$ of the manifold $M$.}
Its graph $(x,v(x))$ represents the {cross--section} of the
tangent bundle $TM$. Velocity vector-fields are cross-sections of the tangent bundle.
Biomechanical \emph{Lagrangian} (that is, kinetic minus potential energy) is a natural energy function on the tangent bundle $TM$. The tangent bundle
is itself a smooth manifold. It has its own tangent bundle, $TTM$. Cross-sections of the second tangent bundle $TTM$ are the acceleration vector-fields.

The left-hand branch of the fundamental covariant force functor $\mathcal{F}_*:TT^*M \to TTM$ depicted in (\ref{covfun}) is Hamiltonian dynamics with its symplectic geometry. It takes place in the {\it cotangent bundle} $T^{\ast }M_{rob}$, defined as follows.
A {\it dual} notion to the tangent space $T_{x}M$ to a smooth manifold
$M$ at a point $x=(x^i)$ with local is its {cotangent space} $T_{x}^{\ast }M$ at
the same point $x$. Similarly to the tangent bundle $TM$, for any smooth $n$D
manifold $M$, there is associated its $2n$D \emph{momentum phase-space manifold}, denoted by $T^{\ast }M$ and called the {\it cotangent bundle}. $T^{\ast }M$
is the disjoint union of all its cotangent spaces $T_{x}^{\ast }M$
at all points $x\in M$, i.e., $T^{\ast }M=\dbigsqcup\limits_{x\in
M}T_{x}^{\ast }M$. Therefore, the cotangent bundle of an
$n-$manifold $M$ is the vector bundle $T^{\ast }M=(TM)^{\ast }$,
the (real) dual of the tangent bundle $TM$. Momentum 1--forms (or, covector-fields) $p_i$ are cross-sections of the cotangent
bundle. Biomechanical \emph{Hamiltonian} (that is, kinetic plus potential energy) is a natural energy function on the
cotangent bundle. The cotangent bundle $T^*M$ is itself a smooth manifold. It has its own tangent bundle, $TT^*M$. Cross-sections of the mixed-second bundle $TT^*M$ are the force 1--forms $F_i= \dot{p}_i$.

There is a unique smooth map from the right-hand branch to the left-hand branch of the diagram (\ref{covfun}): $$TM\ni(x^i,v^i)\mapsto (x^i,p^i)\in T^{\ast }M.$$ It is called the \emph{Legendre transformation}, or \emph{fiber derivative} (for details see, e.g.  \cite{GaneshSprBig,GaneshADG}).

The fundamental covariant force functor $\mathcal{F}_*:TT^*M \to
TTM$ states that the force 1--form $F_i= \dot{p}_i$,
defined on the mixed tangent--cotangent bundle $TT^*M$, causes the
acceleration vector-field $a^i= \dot{\bar{v}}^i$, defined on the
second tangent bundle $TTM$ of the configuration manifold $M$.
The corresponding \textit{contravariant acceleration functor} is
defined as its inverse map, $\mathcal{F}^*:TTM\to TT^*M$.

Representation of human motion is rigorously defined
in terms of {Euclidean} $SE(3)$--groups\footnote{Briefly, the Euclidean SE(3)--group is defined as a semidirect
(noncommutative) product (denoted by $\rhd$) of 3D rotations and 3D translations: ~$%
SE(3):=SO(3)\rhd \mathbb{R}^{3}$. Its most important subgroups are the
following:\\
{{\frame{$%
\begin{array}{cc}
\mathbf{Subgroup} & \mathbf{Definition} \\ \hline
\begin{array}{c}
SO(3),\text{ group of rotations} \\
\text{in 3D (a spherical joint)}%
\end{array}
&
\begin{array}{c}
\text{Set of all proper orthogonal } \\
3\times 3-\text{rotational matrices}%
\end{array}
\\ \hline
\begin{array}{c}
SE(2),\text{ special Euclidean group} \\
\text{in 2D (all planar motions)}%
\end{array}
&
\begin{array}{c}
\text{Set of all }3\times 3-\text{matrices:} \\
\left[
\begin{array}{ccc}
\cos \theta & \sin \theta & r_{x} \\
-\sin \theta & \cos \theta & r_{y} \\
0 & 0 & 1%
\end{array}%
\right]%
\end{array}
\\ \hline
\begin{array}{c}
SO(2),\text{ group of rotations in 2D} \\
\text{subgroup of }SE(2)\text{--group} \\
\text{(a revolute joint)}%
\end{array}
&
\begin{array}{c}
\text{Set of all proper orthogonal } \\
2\times 2-\text{rotational matrices} \\
\text{ included in }SE(2)-\text{group}%
\end{array}
\\ \hline
\begin{array}{c}
\mathbb{R}^{3},\text{ group of translations in 3D} \\
\text{(all spatial displacements)}%
\end{array}
& \text{Euclidean 3D vector space}%
\end{array}%
$}}}} of full rigid--body motion in all main human joints \cite{TijIJHR}. The configuration manifold $M$
for human musculo-skeletal dynamics is defined as a Cartesian product of all included
constrained $SE(3)$ groups, $M=\prod_{j}SE(3)^{j}$ where $j$ labels the active joints. The configuration manifold $M$ is coordinated by local joint coordinates $x^i(t),~i=1,...,n=$ total number of active DOF. The corresponding joint velocities $\dot{x}^i(t)$ live in the \emph{velocity phase space} $TM$, which is the \emph{tangent bundle} of the configuration manifold $M$.

The velocity phase-space $TM$ has the Riemannian geometry with the \textit{local metric form}:
\begin{equation}
\langle g\rangle\equiv ds^{2}=g_{ij}dx^{i}dx^{j}, \label{ddg}
\end{equation}
where $g_{ij}(x)$ is the material metric tensor defined by the biomechanical system's \emph{mass-inertia matrix} and $dx^{i}$
are differentials of the local joint coordinates $x^i$ on $M$. Besides giving the local
distances between the points on the manifold
$M,$ the Riemannian metric form $\langle g\rangle$
defines the system's kinetic energy: $$T=\frac{1}{2}g_{ij}\dot{x}^{i}\dot{x}^{j},$$
giving the \emph{Lagrangian equations} of the conservative skeleton motion with kinetic-minus-potential energy Lagrangian $L=T-V$, with the corresponding \emph{geodesic form}
\begin{equation}
\frac{d}{dt}L_{\dot{x}^{i}}-L_{x^{i}}=0\qquad\text{or}\qquad \ddot{x}^i+\Gamma _{jk}^{i}\dot{x}^{j}\dot{x}^{k}=0, \label{geodes}
\end{equation}%
where subscripts denote partial derivatives, while $\Gamma _{jk}^{i}$ are the Christoffel symbols of
the affine Levi-Civita connection of the biomechanical manifold $M$.

The corresponding momentum phase-space $P=T^*M$ provides a natural \emph{symplectic
structure} that can be defined as follows. As the biomechanical configuration space
$M$ is a smooth $n-$manifold, we can
pick local coordinates $\{dx^{1},...,dx^{n}\}\in M$. Then $%
\{dx^{1},...,dx^{n}\}$ defines a basis of the cotangent space $T_{x}^*M$%
, and by writing $\theta \in T_{x}^*M$ as $\theta
=p_{i}dx^{i}$, we get local coordinates
$\{x^{1},...,x^{n},p_{1},...,p_{n}\}$ on $T^*M$. We can now define the
canonical symplectic form $\omega $ on $P=T^*M$ as:
\begin{equation*}
\omega =dp_{i}\wedge dx^{i},
\end{equation*}
where `$
\wedge $' denotes the wedge or exterior product of exterior differential forms.\footnote{Recall that an \emph{exterior differential form} $\al$ of order $p$ (or, a $p-$\emph{form} $\al$) on a base manifold $X$ is a section of the exterior product bundle $\op\w^p T^*X\to X$. It has the following expression in local coordinates on $X$
$$
\al =\al_{\la_1\dots\la_p}
dx^{\la_1}\w\cdots\w dx^{\la_p} \qquad ({\rm such~that~~}|\al|=p),
$$
where summation is performed over all ordered collections $(\la_1,...,\la_p)$.
$\cO^p(X)$ is the vector space of
$p-$forms on a biomechanical
manifold $X$. In particular, the 1--forms
are called the {\it Pfaffian forms}.}
This $2-$form $\omega $ is obviously independent of the choice of
coordinates $\{x^{1},...,x^{n}\}$ and independent of the base point $%
\{x^{1},...,x^{n},p_{1},...,p_{n}\}\in T_{x}^*M$. Therefore,
it is locally constant, and so $d\omega =0$.\footnote{The canonical $1-$form $\theta $ on $T^*M$ is the unique
$1-$form
with the property that, for any $1-$form $\beta $ which is a section of $%
T^*M$ we have $\beta ^*\theta =\theta $.\par Let $f:M\rightarrow M$ be a diffeomorphism. Then $T^*f$
preserves the canonical $1-$form $\theta $ on $T^*M$, i.e.,
$(T^*f)^*\theta =\theta $. Thus $T^*f$ is
symplectic diffeomorphism.}

If $(P,\omega )$ is a $2n$D symplectic manifold then about each point $%
x\in P$ there are local coordinates
$\{x^{1},...,x^{n},p_{1},...,p_{n}\}$ such that $\omega
=dp_{i}\wedge dx^{i}$. These coordinates are called canonical or
symplectic. By the Darboux theorem, $\omega $ is constant in this
local chart, i.e., $d\omega =0$.

\section{Autonomous Hamiltonian Biomechanics}

We develop autonomous Hamiltonian biomechanics on the configuration
biomechanical manifold $M$ in three steps, following the standard
symplectic geometry prescription (see \cite{GaneshSprSml,GaneshSprBig,GaneshADG,Complexity}):\\

\noindent \textbf{Step A} Find a symplectic \emph{momentum phase--space} $%
(P,\omega )$.

Recall that a symplectic structure on a smooth manifold $M$ is a
nondegenerate closed\footnote{A $p-$form $%
\beta $ on a smooth manifold $M$ is called \textit{closed} if its exterior derivative
$d=\partial_i dx^i$ is equal to zero,
\begin{equation*}
d\beta=0.
\end{equation*}
From this condition one can see that the closed form (the \textit{kernel} of
the exterior derivative operator $d$) is conserved quantity. Therefore,
closed $p-$forms possess certain invariant properties, physically
corresponding to the \emph{conservation laws}.\par
Also, a $p-$form $\beta$ that is an exterior derivative of some $(p-1)-$form
$\alpha$,
\begin{equation*}
\beta=d\alpha,
\end{equation*}
is called \textit{exact} (the \textit{image} of the exterior
derivative operator $d$). By \emph{Poincar\'e lemma,} exact forms prove
to be closed automatically,
\begin{equation*}
d\beta=d(d\alpha)=0.
\end{equation*}\par
Since $d^{2}=0$, \emph{every exact form is closed.} The converse
is only partially true, by Poincar\'{e} lemma: every closed form
is \textit{locally exact}.\par
Technically, this means that given a closed $p-$form $\alpha \in
\Omega ^{p}(U)$, defined on an open set $U$ of a smooth manifold
$M$ any point $m\in U$ has a
neighborhood on which there exists a $(p-1)-$form $\beta \in
\Omega ^{p-1}(U)$ such that $d\beta =\alpha |_{U}.$
In particular, there is a Poincar\'{e} lemma for contractible
manifolds: Any closed form on a smoothly contractible manifold is
exact.} $2-$form $\omega $ on $M$, i.e., for each $x\in M$, $%
\omega (x)$ is nondegenerate, and $d\omega =0$.

Let $T_{x}^*M$ be a cotangent space to $M$ at $m$. The
cotangent bundle $T^*M$ represents a union $\cup _{m\in
M}T_{x}^*M$, together with the standard topology on $T^*M$ and a natural smooth manifold structure, the dimension of
which is twice the dimension of $M$. A $1-$form $\theta $ on $M$
represents a section $\theta :M\rightarrow T^*M$ of the
cotangent bundle $T^*M$.

$P=T^*M$ is our momentum phase--space. On $P$ there is a
nondegenerate
symplectic $2-$form $\omega $ is defined in local joint coordinates $%
x^{i},p_{i}\in U$, $U$ open in $P$, as $\omega =dx^{i}\wedge dp_{i}$. In that case the
coordinates $x^{i},p_{i}\in U$ are called canonical. In a usual
procedure the canonical $1-$form $\theta $ is first defined as
$\theta =p_{i}dx^{i}$, and then the canonical 2--form $\omega $ is
defined as $\omega =-d\theta $.

A \emph{symplectic phase--space manifold} is a pair $(P,\omega )$.\\

\noindent \textbf{Step B} Find a \emph{Hamiltonian vector-field} $X_H$ on $%
(P,\omega)$.

Let $(P,\omega)$ be a symplectic manifold. A vector-field
$X:P\rightarrow TP$ is called \emph{Hamiltonian} if there is a
smooth function $F:P\to\mathbb{R}$ such that $i_X\omega\,=\,dF$
($i_X\omega$ denotes the \emph{interior
product} or \emph{contraction} of the vector-field $X$ and the 2--form $%
\omega$). $X$ is \emph{locally Hamiltonian} if $i_X\omega$ is
closed.

Let the smooth real--valued \emph{Hamiltonian function}
$H:P\rightarrow \mathbb{R}$, representing the total biomechanical
energy $H(x,p)\,=\,T(p)\,+\,V(x)$ ($T$ and $V$ denote kinetic and
potential energy of the system, respectively), be given in local canonical coordinates $%
x^{i},p_{i}\in U$, $U $ open in $P$. The \emph{Hamiltonian vector-field} $%
X_{H}$, condition by $i_{X_{H}}\omega \,=\,dH$, is actually
defined via symplectic matrix $J$, in a local chart $U$, as
\begin{equation}
X_{H}=J\nabla H=\left( \partial _{p_{i}}H,-\partial
_{x^{i}}H\right) ,\qquad J={\small\left(
\begin{matrix}
0 & I \\
-I & 0%
\end{matrix}%
\right)} , \label{HamVec}
\end{equation}
where $I$ denotes the $n\times n$ identity matrix and $\nabla $ is
the gradient operator.\\

\noindent \textbf{Step C} Find a \emph{Hamiltonian phase--flow}
$\phi _{t}$ of $X_{H}$.

Let $(P,\omega )$ be a symplectic phase--space manifold and $%
X_{H}\,=\,J\nabla H$ a Hamiltonian vector-field corresponding to
a smooth real--valued Hamiltonian function $H:P\rightarrow
\mathbb{R}$, on it. If a unique one--parameter group of
diffeomorphisms $\phi _{t}:P\rightarrow P$ exists so that
$\frac{d}{dt}|_{t=0}\,\phi _{t}x=J\nabla H(x)$, it is called the
\emph{Hamiltonian phase--flow}.

A smooth curve $t\mapsto \left( x^{i}(t),\,p_{i}(t)\right) $ on
$(P,\omega )$
represents an \emph{integral curve} of the Hamiltonian vector-field $%
X_{H}=J\nabla H$, if in the local canonical coordinates $x^{i},p_{i}\in U$, $%
U$ open in $P$, \emph{Hamiltonian canonical equations} hold:
\begin{equation}
\dot{q}^{i}=\partial_{p_{i}} H,\qquad
\dot{p}_{i}=-\partial_{q^{i}} H.  \label{HamEq}
\end{equation}

An integral curve is said to be \emph{maximal} if it is not a
restriction of an integral curve defined on a larger interval of
$\mathbb{R}$. It follows from the standard theorem on the
\emph{existence} and \emph{uniqueness} of the solution of a system
of ODEs with smooth r.h.s, that if the manifold $(P,\omega )$ is Hausdorff, then for any point $%
x=(x^{i},p_{i})\in U$, $U$ open in $P$, there exists a maximal
integral curve of $X_{H}\,=\,J\nabla H$, passing for $t=0$,
through point $x$. In case $X_{H}$ is complete, i.e., $X_{H}$ is
$C^{p}$ and $(P,\omega )$ is compact, the maximal integral curve
of $X_{H}$ is the Hamiltonian phase--flow $\phi _{t}:U\rightarrow
U$.

The phase--flow $\phi _{t}$ is \emph{symplectic} if $\omega $ is
constant along $\phi _{t}$, i.e., $\phi _{t}^*\omega =\omega$

($\phi _{t}^*\omega $ denotes the \emph{pull--back}\footnote{Given a map $f:X\to X'$ between the two manifolds, the \emph{pullback} on $X$
of a form $\al$ on $X'$ by $f$ is denoted by $f^*\al$.
The pullback satisfies the relations
\be
f^*(\al\w\bt) =f^*\al\w f^*\bt, \qquad  df^*\al =f^*(d\al),
\ee
for any two forms $\al,\bt\in\cO^p(X)$.} of $\omega $ by $%
\phi _{t}$),

iff ${\frak L}_{X_{H}}\omega \,=0$

(${\frak L}_{X_{H}}\omega $ denotes the \emph{Lie derivative}\footnote{The {\it Lie derivative} ${\frak L}_u\al$ of $p-$form $\al$ along
a vector-field $u$ is defined by Cartan's `magic' formula (see \cite{GaneshSprBig,GaneshADG}):
$$\bL_u\al =u\rfloor d\al +d(u\rfloor\al).
$$
It satisfies the \emph{Leibnitz relation}
$$
\bL_u(\al\w\bt)= \bL_u\al\w\bt +\al\w\bL_u\bt.
$$
Here, the \emph{contraction} $\rfloor$ of a vector-field $u = u^\m\dr_\m$ and a $p-$form $\al =\al_{\la_1\dots\la_p}
dx^{\la_1}\w\cdots\w dx^{\la_p}$
on a biomechanical manifold $X$ is given in local coordinates on $X$ by
$$
u\rfloor\al = u^\m
\al_{\m\la_1\ldots\la_{p-1}}
dx^{\la_1}\w\cdots \w dx^{\la_{p-1}}.
$$
It satisfies the following relation
$$
u\rfloor(\al\w\bt)= u\rfloor\al\w\bt +(-1)^{|\al|}\al\w u\rfloor\bt.
$$} of
$\omega $ upon $X_{H}$).

Symplectic phase--flow $\phi _{t}$ consists of canonical transformations on $%
(P,\omega )$, i.e., diffeomorphisms in canonical coordinates
$x^{i},p_{i}\in U$, $U$ open on all $(P,\omega )$ which leave
$\omega $ invariant. In this case the \emph{Liouville theorem} is
valid: $\phi _{t}$ \emph{preserves}
the \emph{phase volume} on $(P,\omega )$. Also, the system's total energy $%
H$ is conserved along $\phi _{t}$, i.e., $H\circ \phi _{t}=\phi
_{t}$.

Recall that the Riemannian metrics $g\,=\,<,>$ on the
configuration manifold $M$ is a positive--definite quadratic form
$g:TM\rightarrow \mathbb{R}$, in local coordinates $x^{i}\in U$,
$U$ open in $M$, given by (\ref{ddg}) above. Given
the metrics $g_{ij}$, the system's Hamiltonian function represents
a momentum $p$--dependent quadratic form $H:T^*M\rightarrow
\mathbb{R}$ -- the system's kinetic energy
$H(p)\,=T(p)=\,\frac{1}{2}<p,p>$,
in local canonical coordinates $x^{i},p_{i}\in U_{p}$, $U_{p}$ open in $%
T^*M$, given by
\begin{equation}
H(p)=\frac{1}{2}g^{ij}(x,m)\,p_{i}p_{j},  \label{dd19}
\end{equation}%
where $g^{ij}(x,m)=g_{ij}^{-1}(x,m)$ denotes the \emph{inverse}
(contravariant) material \emph{metric tensor}
\begin{equation*}
g^{ij}(x,m)\,=\sum_{\chi =1}^{n}m_{\chi }\delta _{rs}\frac{\partial x^{i}}{%
\partial x^{r}}\frac{\partial x^{j}}{\partial x^{s}}.
\end{equation*}

$T^*M$ is an \emph{orientable} manifold, admitting the
standard \emph{volume form}
\begin{equation*}
\Omega _{\omega
_{H}}=\,{\frac{{(-1)^{{\frac{{N(N+1)}}{{2}}}}}}{{N!}}}\omega
_{H}^{N}.
\end{equation*}

For Hamiltonian vector-field, $X_{H}$ on $M$, there is a base
integral curve $\gamma _{0}(t)\,=\,\left(
x^{i}(t),\,p_{i}(t)\right) $ iff $\gamma _{0}(t)$ is a
\emph{geodesic}, given by the one--form \emph{force equation}
\begin{equation}\label{ddmom}
\dot{\bar{p}}_{i}\equiv \dot{p}_{i}+\Gamma
_{jk}^{i}\,g^{jl}g^{km}\,p_{l}p_{m}=0,\qquad \text{with \qquad }\dot{x}%
^{k}=g^{ki}p_{i}.
\end{equation}

The l.h.s $\dot{\bar{p}}_{i}$ of the covariant momentum equation
(\ref{ddmom})\ represents the {intrinsic} or {Bianchi
covariant derivative} of the momentum with respect to time $t$. Basic relation $\dot{%
\bar{p}}_{i}\,=\,0$ defines the \textit{parallel transport} on
$T^{N}$, the simplest form of human--motion dynamics.
In that case Hamiltonian vector-field $%
X_{H}$ is called the \emph{geodesic spray} and its phase--flow is
called the \emph{geodesic flow}.

For Earthly dynamics in the gravitational \emph{potential} field $%
V:M\rightarrow \mathbb{R}$, the Hamiltonian $H:T^*M\rightarrow \mathbb{%
R}$ (\ref{dd19}) extends into potential form
\begin{equation*}
H(p,x)=\frac{1}{2}g^{ij}p_{i}p_{j}+V(x),
\end{equation*}%
with Hamiltonian vector-field $X_{H}\,=\,J\nabla H$ still defined
by canonical equations (\ref{HamEq}).

A general form of a \emph{driven}, non--conservative Hamiltonian
equations reads:
\begin{equation}
\dot{x}^{i}=\partial _{p_{i}}H,\qquad \dot{p}_{i}=F_{i}-\partial
_{x^{i}}H, \label{FHam}
\end{equation}
where $F_{i}=F_{i}(t,x,p)$ represent any kind of joint--driving
\emph{covariant torques}, including active neuro--muscular--like
controls, as functions of time, angles and momenta, as well as
passive dissipative and elastic joint torques. In the covariant
momentum formulation (\ref{ddmom}), the non--conservative
Hamiltonian equations (\ref{FHam}) become
\begin{equation*}
\dot{\bar{p}}_{i}\equiv \dot{p}_{i}+\Gamma
_{jk}^{i}\,g^{jl}g^{km}\,p_{l}p_{m}=F_{i},\qquad \text{with}\qquad \dot{x}%
^{k}=g^{ki}p_{i}.
\end{equation*}

The general form of autonomous Hamiltonian
biomechanics is given by dissipative, driven Hamiltonian equations on $T^*M$:
\begin{eqnarray}
\dot{x}^{i} &=&\frac{\partial H}{\partial p_{i}}+\frac{\partial
R}{\partial
p_{i}},  \label{fh1} \\
\dot{p}_{i} &=&F_{i}-\frac{\partial H}{\partial x^{i}}+\frac{\partial R}{%
\partial x^{i}},  \label{fh2} \\
x^{i}(0) &=&x_{0}^{i},\qquad p_{i}(0)=p_{i}^{0},  \label{fh3}
\end{eqnarray}%
including \textit{contravariant equation} (\ref{fh1}) -- the
\textit{velocity vector-field}, and \textit{covariant equation}
(\ref{fh2}) -- the \textit{force 1--form} (field), together with initial
joint angles and momenta (\ref{fh3}). Here
$R=R(x,p)$ denotes the Raileigh nonlinear (biquadratic)
dissipation function, and $F_{i}=F_{i}(t,x,p)$ are covariant
driving torques of \textit{equivalent muscular actuators},
resembling muscular excitation and contraction dynamics in
rotational form. The velocity vector-field (\ref{fh1}) and the force $1-$form
(\ref{fh2}) together define the generalized Hamiltonian
vector-field $X_{H}$; the Hamiltonian energy
function $H=H(x,p)$ is its generating function.

As a Lie group, the biomechanical configuration manifold $M=\prod_{j}SE(3)^{j}$ is Hausdorff.\footnote{That is, for every pair of points $%
x_{1},x_{2}\in M$, there are disjoint open subsets (charts)
$U_{1},U_{2}\subset M$ such that $x_{1}\in U_{1}$ and $x_{2}\in
U_{2}$.} Therefore, for $x\,=\,(x^{i},\,p_{i})\in U_{p}$%
, where $U_{p}$ is an open coordinate chart in $T^*M$, there exists a unique
one--parameter group of diffeomorphisms $\phi_t:T^*M\rightarrow T^*M$, that is the \emph{autonomous Hamiltonian phase--flow:}
\begin{eqnarray}\label{fh4}
\phi_t &:&T^*M\rightarrow T^*M:(p(0),x(0))\mapsto (p(t),x(t)),   \\
(\phi_t &\circ & \phi _s\, =\,\phi_{t+s},\quad \phi_0\,=\,\text{identity}),
\notag
\end{eqnarray}
given by (\ref{fh1}--\ref{fh3}) such that
\begin{equation*}
{\frac{{d}}{{dt}}}\left\vert _{t=0}\right. \phi_tx\,=\,J\nabla H(x).
\end{equation*}

\section{Time--Dependent Hamiltonian Biomechanics}

In this section we develop time-dependent Hamiltonian biomechanics.
For this, we first need to extend our autonomous Hamiltonian machinery, using the general concepts of bundles, jets and connections.

\subsection{Biomechanical Bundle}

While standard autonomous Lagrangian biomechanics is developed on the configuration manifold $X$, the \emph{time--dependent
biomechanics} necessarily includes also the real time axis $\mathbb{R}$, so we have an \emph{extended configuration manifold} $\mathbb{R}\times X$. Slightly more generally, the fundamental geometrical structure is the so-called \emph{configuration bundle}
$\pi:X\rightarrow \mathbb{R}$. Time-dependent biomechanics is thus formally developed either on the \emph{extended configuration manifold} $\mathbb{R}\times X$, or on the configuration bundle $\pi:X\rightarrow \mathbb{R}$, using the concept of \textit{jets}, which are based on the idea of \textit{higher--order tangency}, or higher--order
contact, at some designated point (i.e., certain anatomical joint) on a biomechanical configuration manifold $X$.

\begin{figure}[h]
 \centerline{\includegraphics[width=11cm]{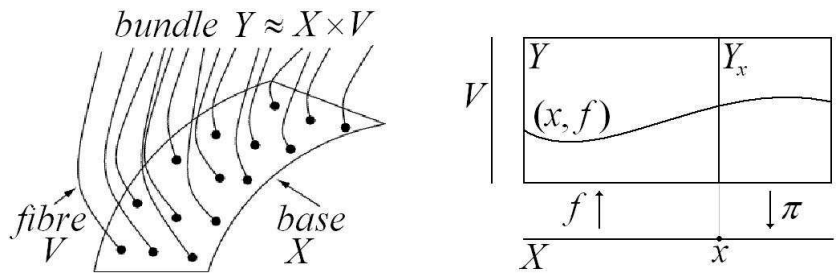}}
\caption{A sketch of a locally trivial fibre bundle $Y\approx X\times V$ as a
generalization of a product space $X\times V$; left -- main
components; right -- a few details (see text for
explanation).}\label{Fibre1}
\end{figure}

In general, tangent and cotangent bundles, $TM$ and $T^{\ast }M$, of a smooth manifold $M$, are special cases of a more general geometrical object called
\emph{fibre bundle}, denoted $\pi
:Y\rightarrow X$, where the word \emph{fiber} $V$ of a map $\pi
:Y\rightarrow X$ is the \emph{preimage} $\pi^{-1}(x)$ of an
element $x\in X$. It is a space which \emph{locally} looks like a
product of two spaces (similarly as a manifold locally looks like
Euclidean space), but may possess a different \emph{global}
structure. To get a visual intuition behind this fundamental
geometrical concept, we can say that a fibre bundle $Y$ is a
\emph{homeomorphic generalization} of a \emph{product space}
$X\times V$ (see Figure \ref{Fibre1}), where $X$ and $V$ are
called the \emph{base} and the \emph{fibre}, respectively. $\pi
:Y\rightarrow X$ is called the \emph{projection}, $Y_{x}=\pi
^{-1}(x)$ denotes a fibre over a point $x$ of the base $X$, while
the map $f=\pi ^{-1}:X\rightarrow Y$ defines the
\emph{cross--section}, producing the \textit{graph} $(x,f(x))$ in
the bundle $Y$ (e.g., in case of a tangent bundle, $f=\dot{x}$
represents a velocity vector--field).

More generally, a biomechanical configuration bundle, $\pi :\Y$, is a locally trivial fibred (or, projection) manifold over the base $X$. It is endowed with an atlas of fibred bundle coordinates $(x^\lambda, y^i)$, where
$(x^\la)$ are coordinates of $X$.

A linear connection $\bar\G$ on a biomechanical bundle $Y\to X$ is given in local coordinates on $Y$ by \cite{book}
\beq
\bar\G=dx^\la\otimes[\dr_\la-\G^i{}_{j\la}(x)y^j\dr_i]. \label{8}
\eeq

An affine connection $\G$ on a biomechanical bundle $Y\to X$ is given in local coordinates on $Y$ by
\be
\G=dx^\la\otimes[\dr_\la+(-\G^i{}_{j\la}(x)y^j+\G^i{}_\la (x)) \dr_i].
\ee
Clearly, a linear connection $\bar\G$ is a special case of an affine connection $\G$.

Every connection $\G$ on a biomechanical bundle $Y\to X$ defines a system of first--order
differential equations on $Y$, given by the local
coordinate relations
\beq
y^i_\la =\G^i(y). \label{39}
\eeq
Integral sections for $\G$ are local solutions of (\ref{39}).

\subsection{Biomechanical Jets}

\begin{figure}[h]
\centerline{\includegraphics[width=6cm]{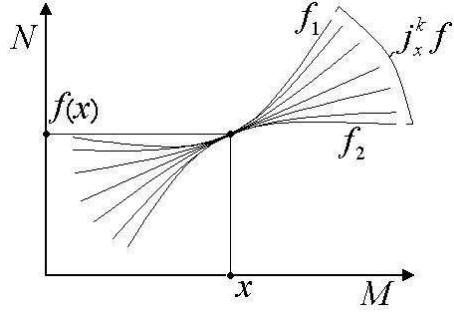}} \caption{An
intuitive geometrical picture behind the $k-$jet concept, based on
the idea of a higher--order tangency (or, higher--order contact). }
\label{jet1}
\end{figure}

A
pair of smooth manifold maps, ~$f_{1},f_{2}:M\rightarrow N$~ (see
Figure \ref{jet1}), are said to be $k-$\emph{tangent} (or
\emph{tangent of order }$k$, or
have a $k$th \emph{order contact}) at a point $x$ on a domain manifold $M$, denoted by $%
f_{1}\sim f_{2}$, iff
\begin{eqnarray*}
f_{1}(x) &=&f_{2}(x)\qquad \text{called}\quad 0-\text{tangent}, \\
\partial _{x}f_{1}(x) &=&\partial _{x}f_{2}(x),\qquad \text{called}\quad 1-%
\text{tangent}, \\
\partial _{xx}f_{1}(x) &=&\partial _{xx}f_{2}(x),\qquad \text{called}\quad 2-%
\text{tangent}, \\
&&...\qquad \text{etc. to the order }k
\end{eqnarray*}
In this way defined $k-$\emph{tangency} is an \emph{equivalence
relation}.

A $k-$\textit{jet} (or, a \emph{jet of order }$k$), denoted
by $j_{x}^{k}f$, of a smooth map $f:M\rightarrow N$ at a
point $x\in M$ (see Figure \ref {jet1}), is defined as an
\emph{equivalence class} of $k-$tangent maps at $x$,
\begin{equation*}
j_{x}^{k}f:M\rightarrow N=\{f':f'\text{ is }k-\text{tangent to
}f\text{ at }x\}.
\end{equation*}

For example, consider a simple function
~$f:X\rightarrow Y,\,x\mapsto y=f(x)$, mapping the $X-$axis
into the $Y-$axis in $\mathbb{R}^2$. At a chosen point
$x\in X$ we have:\\ a $0-$jet is a graph: $(x,f(x))$;\\
a $1-$jet is a triple: $(x,f(x),f{'}(x))$;\\ a
$2-$jet is a quadruple: $(x,f(x),f{'}(x),f^{\prime \prime }(x))$,\\
~~ and so on, up to the order $k$ (where
$f{'}(x)=\frac{df(x)}{dx}$, etc).\\ The set of all $k-$jets from
$j^k_xf:X\rightarrow Y$ is called the $k-$jet manifold
$J^{k}(X,Y)$.

Formally, given a biomechanical bundle $\Y$, its first--order {\it jet manifold} $J^1Y$ comprises the set of
equivalence classes $j^1_xs$, $x\in X$, of sections $s:X\to Y$
so that sections
$s$ and $s'$ belong to the same class iff
$$
Ts\mid _{T_xX} =Ts'\mid_{T_xX}.
$$
Intuitively, sections $s,s'\in j^1_xs$  are identified by their values
$s^i(x)={s'}^i(x)$ and the values of their partial derivatives
$\dr_\mu s^i(x)=\dr_\mu{s'}^i(x)$
at the point $x$ of $X$. There are the natural fibrations \cite{book}
$$
\pi_1:J^1Y\ni j^1_xs\mapsto x\in X, \qquad
\pi_{01}:J^1Y\ni j^1_xs\mapsto s(x)\in Y.
$$
Given bundle coordinates $(x^\la,y^i)$ of $Y$, the associated jet manifold $J^1Y$
is endowed with the adapted coordinates
\be
(x^\la,y^i,y_\la^i), \qquad (y^i,y_\la^i)(j^1_xs)=(s^i(x),\dr_\la
s^i(x)), \qquad
{y'}^i_\la = \frac{\dr x^\m}{\dr{x'}^\la}(\dr_\m +y^j_\m\dr_j)y'^i.
\ee

In particular, given the biomechanical configuration bundle $M\rightarrow \mathbb{R}$ over
the time axis $\mathbb{R}$, the \textit{$1-$jet
space} $J^{1}(\mathbb{R},M)$
is the set of equivalence classes $j_{t}^{1}s$ of sections $s^{i}:\mathbb{R}%
\rightarrow M$ of the configuration bundle $M\rightarrow \mathbb{R}$, which are
identified by their values $s^{i}(t)$, as well as by the values of their partial derivatives $%
\partial _{t}s^{i}=\partial _{t}s^{i}(t)$ at time points $t\in \mathbb{R}$.
The 1--jet manifold $J^{1}(\mathbb{R},M)$ is coordinated by $(t,x^{i},\dot{x}%
^{i})$, that is by \textsl{(time, coordinates and velocities)} at every active human joint, so the 1--jets are local joint coordinate maps
\begin{equation*}j_{t}^{1}s:\mathbb{R}%
\rightarrow M,\qquad t\mapsto (t,x^{i},\dot{x}^{i}).
\end{equation*}

The {\it repeated jet manifold}
$J^1J^1Y$  is defined to be the jet manifold of the bundle
$J^1Y\to X$. It
is endowed with the adapted coordinates $(x^\la ,y^i,y^i_\la
,y_{(\m)}^i,y^i_{\la\m})$.

The {\it second--order jet manifold} $J^2Y$
of a bundle $\Y$ is the subbundle of $\wh J^2Y\to J^1Y$ defined
by the coordinate conditions
$y^i_{\la\m}=y^i_{\m\la}$. It has the local coordinates
$(x^\la ,y^i, y^i_\la,y^i_{\la\leq\m})$
together with the transition functions \cite{book}
\be
{y'}_{\la\m}^i= \frac{\dr x^\al}{\dr{x'}^\m}(\dr_\al +y^j_\al\dr_j
+y^j_{\nu\al}\dr^\nu_j){y'}^i_\la.
\ee
The second--order jet manifold $J^2Y$  of $Y$ comprises
the equivalence classes  $j_x^2s$ of sections $s$ of\\ $Y\to X$ such that
\be
y^i_\la (j_x^2s)=\dr_\la s^i(x),\qquad
y^i_{\la\m}(j_x^2s)=\dr_\m\dr_\la s^i(x).
\ee
In other words, two sections $s,s'\in j^2_xs$ are identified by their values
and the values of their first and second--order
derivatives at the point $x\in X$.

In particular, given the biomechanical configuration bundle $M\rightarrow \mathbb{R}$ over
the time axis $\mathbb{R}$, the \textit{$2-$jet space} $J^{2}(\mathbb{R},M)$
is the set of equivalence classes $j_{t}^{2}s$ of sections $s^{i}:\mathbb{R}\rightarrow M$%
\ of the configuration bundle $\pi:M\rightarrow \mathbb{R}$, which
are identified by their values $s^{i}(t)$, as well as the values
of their first and second partial derivatives, $\partial
_{t}s^{i}=\partial _{t}s^{i}(t)$
and $\partial _{tt}s^{i}=\partial _{tt}s^{i}(t)$, respectively, at time points $%
t\in \mathbb{R}$. The 2--jet manifold $J^{2}(\mathbb{R},M)$ is
coordinated by $(t,x^{i},\dot{x}^{i},\ddot{x}^{i})$, that is by \textsl{(time, coordinates, velocities and accelerations)} at every active human joint, so the
2--jets are local joint coordinate maps\footnote{For more technical details on jet spaces with their physical applications, see \cite{book,sard98}).}
\begin{equation*}j_{t}^{2}s:\mathbb{R}%
\rightarrow M,\qquad t\mapsto
(t,x^{i},\dot{x}^{i},\ddot{x}^{i}).
\end{equation*}

\subsection{Polysymplectic Dynamics}

Let $Y\to X$ be a biomechanical bundle with local coordinates
$(x^\la, y^i)$. In jet terms, a \textit{first--order Lagrangian} is defined (through the standard Lagrangian density $\cL$) to
be a horizontal density $L=\cL\om$
on the jet manifold $J^1Y$.
The jet manifold $J^1Y$ plays the role of the
finite-dimensional configuration space of
sections of the bundle $Y\to X$.

Lagrangian $L$ yields the \textit{Legendre morphism}
$\wh L:J^1Y\to\Pi$, from the 1--jet manifold $J^1Y$ to the \textit{Legendre manifold}, given by a product \cite{book}
\beq
\Pi=V^*Y\op\w(\op\w^{n-1} T^*X)  =V^*Y\op\w(\op\w^n
T^*X) \op\otimes TX,  \label{00}
\eeq
where $V$ is called the \emph{vertical bundle}.
$\Pi$ plays the role of the finite-dimensional phase-space of sections of $\Y$.
Given the bundle coordinates $(x^\la, y^i)$ on $\Y$,
the Legendre bundle (\ref{00}) has local coordinates
$(x^\la ,y^i,p^\la_i)$, where $p^\la_i$ are the holonomic coordinates with
the transition functions
\beq
{p'}^\la_i = \det (\frac{\dr x^\ve}{\dr {x'}^\nu}) \frac{\dr y^j}{\dr{y'}^i}
\frac{\dr {x'}^\la}{\dr x^\m}p^\m_j.  \label{2.3}
\eeq
Relative to these coordinates, the Legendre morphism  $\wh L$ reads
\beq
 p^\m_i\circ\wh L=\pi^\m_i. \label{m11}
\eeq

The Legendre manifold $\Pi$ in (\ref{00}) is equipped with
the \textit{generalized Liouville form}
\beq
\bth =-p^\la_idy^i\w\om\otimes\dr_\la, \label{2.4}
\eeq
where $\otimes$ denotes the standard tensor product. Furthermore, for any Pfaffian form $\theta$ on $X$ we have the relation
\be
\bla\rfloor\theta = -d(\bth\rfloor\theta).
\ee
This relation introduces the the \textit{canonical polysymplectic form} on the Legendre manifold $\Pi$,
\beq
\bla =dp^\la_i\w dy^i\w\om\otimes\dr_\la,  \label{406}
\eeq
whose coordinate expression (\ref{406}) is maintained under holonomic
coordinate transformations (\ref{2.3}). It is a pullback-valued form \cite{book}.

\subsection{Hamiltonian Connections}

Let $J^1\Pi$ be the jet manifold of the \emph{Legendre bundle}
$\Pi\to X$. It is endowed with the adapted  coordinates
$( x^\la, y^i, p^\la_i, y^i_\m, p^\la_{i\m})$.
By analogy with the notion of an autonomous Hamiltonian
vector-field $X_H$ in (\ref{HamVec}), a connection $\g$ on the bundle $\Pi\to X$, given by
\be
\g =dx^\la\otimes(\dr_\la +\g^i_\la\dr_i +\g^\m_{i\la}\dr^i_\m),
\ee
is said to be \textit{locally
Hamiltonian} if the exterior form
$\g\rfloor\bla$ is closed and \textit{Hamiltonian} if the form
$\g\rfloor\bla$ is exact. A connection $\g$ is locally Hamiltonian
iff it obeys the conditions \cite{book,sard98}
\beq
\dr^i_\la\g^j_\m-\dr^j_\m\g^i_\la=0,\quad
 \dr_i\g_{j\m}^\m- \dr_j\g_{i\m}^\m=0,\quad
 \dr_j\g_\la^i+\dr_\la^i\g_{j\m}^\m=0.\label{422}
\eeq

A $p-$form $H$ on the
Legendre bundle $\Pi\to X$ is called a
\textit{general Hamiltonian form} if
there exists a Hamiltonian connection such that
\be
\g\rfloor\bla = dH.
\ee
A general, dissipative, time-dependent Hamiltonian form $H$ on $\Pi$
is said to be \textit{Hamiltonian} if it
has the splitting
\beq
H=H_\G -\wt H_\G=p^\la_idy^i\w\om_\la -
(p^\la_i\G^i_\la +\wt{\cH}_\G)\om
 =p^\la_idy^i\w\om_\la-\cH\om  \label{4.7}
\eeq
modulo closed forms, where $\G$ is a
connection on $Y$ and $\wt H_\G$ is a horizontal density.
This splitting is preserved under the holonomic coordinate
transformations (\ref{2.3}).

Let a Hamiltonian connection $\g$ associated with a Hamiltonian form $H$ have
an integral section $s$ of
$\Pi\to X$, that is, $\g\circ s=J^1s$. Then $s$ satisfies the
system of first--order \textit{Hamiltonian equations} on $\Pi$,
$$
y^i_\la =\dr^i_\la\cH, \qquad
p^\la_{i\la} =-\dr_i\cH.
$$

\subsection{Time--Dependent Dissipative Hamiltonian Dynamics}

We can now formulate the time-dependent biomechanics as an $n=1-$reduction of polysymplectic dynamics described above. The biomechanical phase space is the Legendre manifold $\Pi$, endowed with
the holonomic coordinates $(t,y^i,p_i)$
with the transition functions
$$
p'_i=\frac{\dr y^j}{\dr {y'}^i}p_j.
$$
$\Pi$ admits
the canonical form $\bla$ (\ref{406}), which now reads
$$
\bla=dp_i\w dy^i\w dt\ot\dr_t.
$$
As a particular case of the polysymplectic machinery of the previous section,
we say that a connection
\be
\g =dt\ot(\dr_t +\g^i\dr_i +\g_i\dr^i)
\ee
on the bundle $\Pi\to X$ is locally
Hamiltonian if the exterior form
$\g\rfloor\bla$ is closed and Hamiltonian if the form $\g\rfloor\bla$ is exact. A connection $\g$ is
locally Hamiltonian iff it obeys the conditions (\ref{422}) which
now take the form
\be
\dr^i\g^j-\dr^j\g^i=0,\quad
\dr_i\g_j- \dr_j\g_i=0,\quad \dr_j\g^i+\dr^i\g_j=0.
\ee

Note that every connection
$\G=dt\ot(\dr_t +\G^i\dr_i)$ on the bundle
$\Y$ gives rise to the Hamiltonian connection $\wt\G$ on $\Pi\to X$, given by
$$
\wt\G =dt\ot(\dr_t +\G^i\dr_i -\dr_j\G^i p_i\dr^j).
$$
The corresponding Hamiltonian form $H_\G$ is given by
$$
H_\G=p_idy^i -p_i\G^idt.
$$

Let $H$ be a dissipative Hamiltonian form (\ref{4.7}) on $\Pi$, which reads:
\beq
H=p_idy^i-\cH dt=p_idy^i -p_i\G^idt -\wt{\cH}_\G dt.  \label{m46}
\eeq
We call $\cH$ and $\wt\cH$ in the decomposition (\ref{m46}) the
\textit{Hamiltonian} and the \textit{Hamiltonian function} respectively.
Let $\g$ be a Hamiltonian connection on $\Pi\to X$ associated with the
Hamiltonian form (\ref{m46}). It satisfies the relations \cite{book,sard98}
\begin{eqnarray}
&&\g\rfloor\bla =dp_i\w dy^i+ \g_idy^i\w dt -\g^idp_i\w dt = dH,\nonumber\\
&&\g^i =\dr^i\cH, \qquad \g_i=-\dr_i\cH. \label{m40}
\end{eqnarray}
From equations (\ref{m40}) we see that, in the case of biomechanics,
one and only one Hamiltonian connection is associated with a given
Hamiltonian form.

Every connection $\g$ on $\Pi\to X$
yields the system of first--order
differential equations (\ref{39}), which now takes the explicit form:
\beq
\dot{y}^i =\g^i, \qquad \dot{p}_i =\g_i.\label{m170}
\eeq
They are called the \textit{evolution equations}.
If $\g$ is a Hamiltonian connection associated with the Hamiltonian form $H$
(\ref{m46}), the evolution equations (\ref{m170}) become the \emph{dissipative time-dependent
Hamiltonian equations}:
\begin{eqnarray}
\dot{y}^i =\dr^i\cH, \qquad
\dot{p}_i =-\dr_i\cH. \label{m41}
\end{eqnarray}

In addition, given any scalar function $f$ on
$\Pi$, we have the \textit{dissipative Hamiltonian evolution equation}
\beq
d_{H}f=(\dr_t +\dr^i\cH\dr_i -\dr_i\cH\dr^i)\,f,
\label{m59}
\eeq
relative to the Hamiltonian $\cH$. On solutions $s$ of the Hamiltonian
equations (\ref{m41}), the evolution equation (\ref{m59}) is equal to the total time
derivative of the function $f$:
\be
s^*d_{H}f=\frac{d}{dt}(f\circ s).
\ee

\subsection{Time\,\&\,Fitness--Dependent Biomechanics}

The dissipative Hamiltonian system (\ref{m41})--(\ref{m59}) is the basis for our time\,\&\,fitness-dependent biomechanics. The scalar function $f$ in (\ref{m59}) on the biomechanical Legendre phase-space manifold $\Pi$ is now interpreted as an \emph{individual neuro-muscular fitness function}. This fitness function is a `determinant' for the performance of muscular drives for the driven, dissipative Hamiltonian biomechanics. These muscular drives, for all active DOF, are given by time\,\&\,fitness-dependent Pfaffian form: $F_i=F_i(t,y,p,f)$. In this way, we obtain our final model for time\,\&\,fitness-dependent Hamiltonian biomechanics:
{\large\begin{eqnarray*}
\dot{y}^i &=& \dr^i\cH, \\
\dot{p}_i &=& F_i-\dr_i\cH, \\
d_{H}f &= &(\dr_t +\dr^i\cH\dr_i -\dr_i\cH\dr^i)\,f.
\end{eqnarray*}}

Physiologically, the active muscular drives $F_i=F_i(t,y,p,f)$ consist of
\cite{GaneshSprSml,GaneshWSc}):\\

\textbf{1. Synovial joint mechanics}, giving the first stabilizing effect
to the conservative skeleton dynamics, is described by the
$(y,\dot{y})$--form of the \textit{Rayleigh--Van der Pol's dissipation
function}
\begin{equation*}
R=\frac{1}{2}\sum_{i=1}^{n}\,(\dot{y}^{i})^{2}\,[\alpha _{i}\,+\,\beta
_{i}(y^{i})^{2}],\quad
\end{equation*}
where $\alpha _{i}$ and $\beta _{i}$ denote dissipation parameters. Its
partial derivatives give rise to the viscous--damping torques and forces in
the joints
\begin{equation*}
\mathcal{F}_{i}^{joint}=\partial R/\partial \dot{y}^{i},
\end{equation*}
which are linear in $\dot{y}^{i}$ and quadratic in $y^{i}$.\\

\textbf{2. Muscular mechanics}, giving the driving torques and forces $%
\mathcal{F}_{i}^{musc}=\mathcal{F}_{i}^{musc}(t,y,\dot{ y})$ with $%
(i=1,\dots ,n)$ for human biomechanics, describes the internal {excitation} and
{contraction} dynamics of {\it equivalent muscular actuators} \cite%
{Hatze}.\\

(a) The \textit{excitation dynamics} can be described by an impulse {%
force--time} relation
\begin{eqnarray*}
F_{i}^{imp} &=&F_{i}^{0}(1\,-\,e^{-t/\tau _{i}})\text{ \qquad if stimulation
}>0 \\
\quad F_{i}^{imp} &=&F_{i}^{0}e^{-t/\tau _{i}}\qquad \qquad \;\quad\text{if
stimulation }=0,\quad
\end{eqnarray*}
where $F_{i}^{0}$ denote the maximal isometric muscular torques
and forces, while $\tau _{i}$ denote the associated time
characteristics of particular muscular actuators. This relation
represents a solution of the Wilkie's muscular {\it active--state
element} equation \cite{Wilkie}
\begin{equation*}
\dot{\mu}\,+\,\G \,\mu \,=\,\G \,S\,A,\quad \mu (0)\,=\,0,\quad
0<S<1,
\end{equation*}
where $\mu =\mu (t)$ represents the active state of the muscle, $\G $
denotes the element gain, $A$ corresponds to the maximum tension the element
can develop, and $S=S(r)$ is the `desired' active state as a function of the
motor unit stimulus rate $r$. This is the basis for biomechanical force controller.\\

(b) The \textit{contraction dynamics} has classically been described by the
Hill's \textit{hyperbolic force--velocity} relation \cite{Hill}
\begin{equation*}
F_{i}^{Hill}\,=\,\frac{\left( F_{i}^{0}b_{i}\,-\,\delta _{ij}a_{i}\dot{y}%
^{j}\,\right) }{\left( \delta _{ij}\dot{y}^{j}\,+\,b_{i}\right) },\,\quad
\end{equation*}
where $a_{i}$ and $b_{i}$ denote the {Hill's parameters},
corresponding to the energy dissipated during the contraction and
the phosphagenic energy conversion rate, respectively, while
$\delta _{ij}$ is the Kronecker's $\delta-$tensor.

In this way, human biomechanics describes the excitation/contraction dynamics for the $i$th
equivalent muscle--joint actuator, using the simple impulse--hyperbolic product relation
\begin{equation*}
\mathcal{F}_{i}^{musc}(t,y,\dot{y})=\,F_{i}^{imp}\times F_{i}^{Hill}.\quad
\end{equation*}

Now, for the purpose of biomedical engineering and rehabilitation,
human biomechanics has developed the so--called \textit{hybrid rotational actuator}. It
includes, along with muscular and viscous forces, the D.C. motor
drives, as used in robotics
\begin{eqnarray*}
&&\mathcal{F}_{k}^{robo}=i_{k}(t)-J_{k}\ddot{y}_{k}(t)-B_{k}\dot{y}_{k}(t),\qquad\text{with}\\
&&l_{k}i_{k}(t)+R_{k}i_{k}(t)+C_{k}\dot{y}_{k}(t)=u_{k}(t),
\end{eqnarray*}
where $k=1,\dots,n$, $i_{k}(t)$ and $u_{k}(t)$ denote currents and voltages
in the rotors of the drives, $R_{k},l_{k}$ and $C_{k}$ are resistances,
inductances and capacitances in the rotors, respectively, while $J_{k}$ and $%
B_{k}$ correspond to inertia moments and viscous dampings of the drives,
respectively.

Finally, to make the model more realistic, we need to add some \textit{stochastic
torques and forces}:
\begin{equation*}
\mathcal{F}_{i}^{stoch}=B_{ij}[y^{i}(t),t]\,dW^{j}(t),
\end{equation*}
where $B_{ij}[y(t),t]$ represents continuous stochastic \textit{diffusion
fluctuations}, and $W^{j}(t)$ is an $N-$variable \textit{Wiener process}
(i.e., generalized Brownian motion) \cite{StrAttr}, with
$$dW^{j}(t)=W^{j}(t+dt)-W^{j}(t),\qquad (\text{for} ~~j=1,\dots,n=\text{no. of active DOF}).$$

\bigbreak
\section{Conclusion}

We have presented the time-dependent generalization of an `ordinary' autonomous Hamiltonian biomechanics, in which total mechanical + biochemical energy is not conserved. Starting with the Covariant Force Law, we have first developed autonomous Hamiltonian biomechanics. Then we have introduced a general framework for time-dependent Hamiltonian biomechanics in terms jets, Legendre manifolds and dissipative Hamiltonian connections associated with the extended musculo-skeletal configuration manifold, called the configuration bundle. In this way we formulated a general Hamiltonian model for time\,\&\,fitness-dependent human biomechanics.

\end{document}